\documentclass[aps,prb,citeautoscript,reprint,showpacs,superscriptaddress,twocolumn]{revtex4-2}

\usepackage{preamble}

\begin{document}

\date{\today}

\title{Majorana Loop Models for Measurement-Only Quantum Circuits}

\author{Kai Klocke}
\affiliation{Department of Physics, University of California, Berkeley, California 94720, USA}
\author{Michael Buchhold}
\affiliation{Institut f\"ur Theoretische Physik, Universit\"at zu K\"oln, D-50937 Cologne, Germany}

\begin{abstract}
Projective measurements in random quantum circuits lead to a rich breadth of entanglement phases and extend the  realm of non-unitary quantum dynamics.
Here we explore the connection between measurement-only quantum circuits in one spatial dimension and the statistical mechanics of loop models in two dimensions.
While Gaussian Majorana circuits admit a microscopic mapping to loop models, for non-Gaussian, i.e., generic Clifford, circuits a corresponding mapping may emerge only on a coarse grained scale.
We then focus on a fundamental symmetry of loop models: the orientability of world lines.
We discuss how orientability enters in the measurement framework, acting as a separatrix for the universal long-wavelength behavior in a circuit.
When orientability is broken, the circuit falls into the universality class of closely packed loops with crossings (CPLC) and features a Goldstone phase with a peculiar, universal $\log^2(L)$-scaling of the entanglement entropy. 
In turn, when orientability is preserved, the long-wavelength behavior of the circuit mimics that of (coupled) two-dimensional Potts models. 
We demonstrate the strength of the loop model approach by numerically simulating a variety of measurement-only Clifford circuits. 
Upon varying the set of measured operators, a rich circuit dynamics is observed, ranging from CPLC to the $1$-state Potts model (percolation), the $2$-state Potts model (Ising) and coupled Potts models (BKT) universality class. 
Loop models thus provide a handle to access a large class of measurement-only circuits and yield a blueprint on how to realize desired entanglement phases by measurement.
\end{abstract}

\maketitle

\section{Introduction}\label{sec:intro}
Quantum circuits in the NISQ era represent an exciting playground to explore novel types of quantum dynamics~\cite{Preskill_NISQ}. 
Recent examples range from testing elementary digital quantum computing algorithms, such as error correction~\cite{Ebadi_2022, Erhard_2021, Postler_2022, Stricker_2020, Nigg_2014, Krinner_2022, Ryan_Anderson_2021} or topological state preparation~\cite{Semeghini_2021, Weber_2022, ippoliti_google_2023, Iqbal_2023_nonabelian, Iqbal_2023_topological, FossFeig_2023}, to the realization of nonequilibrium quantum many-body states without an equilibrium counterpart~\cite{Noel_2022, Cherktov_2022}.

A particular role, unique to quantum circuits, is played by measurements. 
Evolution generated by measurement yields irreversible, non-unitary quantum dynamics, which provides \emph{directionality} to a circuit. 
At the same time, measurements preserve the purity of a wave function. 
This enables a genuine quantum evolution, based only on the non-commutativity of operators, i.e., of the generators of the dynamics. 
Applying sequences of consecutive measurements thus provides the potential to implement new types of directional quantum dynamics in order to evolve and to prepare desired quantum many-body states in near-term NISQ devices. 
This prospect has lead to fruitful activity amongst a broad, interdisciplinary research community. 
A prominent example for the novel type of quantum dynamics induced by measurement are measurement-induced phase transitions in the behavior of the entanglement entropy~\cite{Fisher_2023_review}. 

Entanglement transitions induced by measurements, either competing with unitary gates~\cite{Turkeshi_2020_2D, Sierant_2022_higherDim, Sharma_2022, Bao_2020, Choi_2020, Block_2022, Weinstein_2022_negativity, Li_2018_zeno, Li_2019, Li_2021_CFT, Li_2021_statmech_qecc, Li_2023_DPRE, Yang_2022, Li_2021_statmech_tensor, Sang_2021, Gullans_2020, Gullans_2020_probe, Oshima_2023, Jian_2020, Jian_2022, Jian_2023, Iaconis_2020, Iaconis_2021}, Hamiltonian evolution~\cite{Altland_2022, Minoguchi_2022, Muller_2022, Buchhold_2021_PRX, Alberton_2021, Buchhold_2022, Zhang_2023, Zhou_2023, Turkeshi_2021, Sierant_2022_floquet, Turkeshi_2022_resetting, Biella_2021, Turkeshi_2022_bdry, Gal_2023, Casagrande_2023, Fuji_2020, Jian_2021_SYK} or other, incommensurate measurements~\cite{Ippoliti_2021, Lang_2020, Sriram_2022, Zhu_2023, Li_2021_Z2, Sang_2021, Hauser_2023, Sharma_2023} have advanced our understanding of non-unitary quantum dynamics and have established new links between research areas. 
Measurement-induced phase transitions have, for instance, been connected to quantum error correction~\cite{Li_2021_statmech_qecc, Li_2021_Z2, Gullans_2020, Choi_2020}, measurement-based quantum computation~\cite{Sharma_2023}, quantum state preparation~\cite{zhu2022,Lu_prep,tantivasadakarn2022longrange} and quantum teleportation~\cite{ippoliti_google_2023, Bao_2022_teleportation}, bridging quantum information science, where measurements traditionally play a vital role, with non-equilibrium physics and aspects of statistical mechanics, e.g., non-unitary conformal field theories.

Here we undertake a closer inspection of one further interdisciplinary theme: the relation between (1+1)-dimensional measurement-only quantum circuits and two-dimensional (2D) loop models.
Loop models have a long and rich history, touching many aspects of physics ranging from the classical statistical mechanics of polymer collapse to quantum knot invariants in topological quantum field theory. 
In the last decade there has been a renewed interest in better understanding universality classes which emerge from loop models subject to particular dynamical constraints or perturbations \cite{Dai_2020, Nahum_2013, Nahum_2013_3D_sigma, Nahum_2013_loopSoup, Nahum_2015, Nahum_2016, Nahum_2017, Fendley_2006, Fendley_2008, Fendley_2008_quantum, Pearce_2007, Vernier_2016}. 
More recently, entanglement dynamics in a particular family of random Clifford circuits has been expressed in terms of Majorana fermion worldlines, making concrete connection to well studied loop models \cite{Sang_2021_MPQP, Nahum_2020, Sang_2021}.
In this work we expand this connection between loop models and random Clifford circuits. 
We argue, based on general considerations, that a large class of both Gaussian and non-Gaussian measurement-only circuits can be understood in an (emergent) loop model framework. 
We confirm this perspective by performing numerical simulations for a set of measurement-only circuits.

First, we show how in one dimension (1D), Gaussian Majorana projection operators and SWAP gates generate either the paradigmatic Temperley-Lieb (TL) or, more generally, the Brauer algebra. 
Consequently, any 1D circuit constructed from discrete, Gaussian operations admits a loop model representation. 
Based on the particular form of the algebra, we introduce a key concept for loop models and the corresponding circuits: the loop orientability.

Orientability is a hidden $U(1)$ symmetry, which determines the long-wavelength behavior in the circuit.
If present, the dynamics in the circuit are related to loop models without crossings, which can be furnished with a TL algebra, such as the 2D Potts model.
When orientability is broken, however, the circuit approaches the long-wavelength behavior of a closely packed loop model with crossings (CPLC)~\cite{Nahum_2013}. 
The latter gives rise to an unconventional symmetry broken \emph{Goldstone} phase with a characteristic $\sim \log^2(L)$ growth of the entanglement entropy. 
We show that in a circuit, this symmetry has an interpretation in terms of both the circuit geometry and the form of the generators of the evolution: 
Firstly, orientable circuits admit a \emph{bipartition} of the underlying Majorana lattice such that only parities between different sublattices are measured.
Secondly, this is equivalent of finding a representation of the circuit where all generators of the evolution obey \emph{time-reversal symmetry}. 

The latter provides an angle to extend the symmetry to non-Gaussian, i.e., generic, random Clifford, circuits.
The measurement of operators that are even under time-reversal preserve orientable structures, while measuring operators that are odd under time-reversal will break orientability. 
As a result, the former provides an interpretation of non-Gaussian measurement-only circuits in terms of coupled, i.e. \emph{interacting}, $Q$-state Potts models, which may admit an exact analytical solution, while the latter pushes the circuit into a non-Gaussian CPLC. 

We confirm this picture by numerically simulating measurement-only Clifford circuits with nearest and next-nearest neighbor qubit measurements. 
Whenever the measurements break orientability, either in the Gaussian or non-Gaussian case, we find that the long wavelength behavior of the circuit is described by the CPLC. 
Depending on the set of performed measurements, the circuit realizes a set of topologically distinct area law phases, which are separated from each other by an extended Goldstone phase. 
The Goldstone phase and its critical line are described by the CPLC nonlinear sigma model~\cite{Nahum_2013}, or an enlarged supersymmetric field theory~\cite{Jacobsen_2003}, yielding universal behavior for the entanglement growth, including the numerical prefactors, which are exactly recovered by the (non-) Gaussian circuits. 

In turn, for the orientable nearest-neighbor circuits, we establish a direct connection to coupled $1$-state Potts models via the transfer matrix method and the Temperley-Lieb algebra. 
This provides an analytical approach to understand the rich phase diagram which arises in the orientable non-Gaussian model. 
Depending on the parameter regime, it realizes a series of measurement-induced phases and phase transitions: 
(i) A phase transition between topologically distinct area law phases, which is described by the $1$-state Potts model (equivalent to 2D bond percolation). 
(ii) The branching of the $1$-state Potts model critical point into two Berezhinskii-Kosterlitz-Thouless (BKT) transitions, separated by an extended critical phase, which is described by two ferromagnetically coupled $1$-state Potts models. 
(iii) An emergent volume law phase, where Majorana worldlines braid freely due to the competition between non-Gaussian measurements.
(iv) A novel measurement-induced phase transition separating two topologically distinct, non-Gaussian area law phases, which is described by the Ising universality class of the $2$-state Potts model. 
We confirm these scenarios by matching the critical exponents $\nu, \eta, \eta_\parallel$ to the known values for  Potts models.

The link to loop models, and the focus on orientability as a relevant symmetry, gives rise to a practical approach both to classify and to analytically solve many classes of measurement-only circuits.
Further, it provides a toolbox to realize desired loop model phases by implementing the corresponding classes of (non-) Gaussian measurements. 

The paper is organized as follows.
We begin in Sec.~\ref{sec:loop_review} by reviewing fundamental aspects of loop models in two spatial dimensions (2D) and by linking them to the dynamics in (1+1)-dimensional measurement-only Clifford circuits.
In Sec.~\ref{sec:loop_circuits} we give a concrete mapping between Gaussian Majorana circuits and 2D loop models, and we discuss the loop model observables in Sec.~\ref{sec:observables}. 
We then introduce the notion of worldline orientability and its role in classifying measurement-only circuits in Sec.~\ref{sec:orientability}.
We conclude this section by examining the robustness of the loop model picture when including non-Gaussian measurements in Secs.~\ref{sec:beyond_ff} and \ref{sec:beyond_ff}.  

In Sec.~\ref{sec:measurement_only_ladders} we examine a set of (non-) Gaussian measurement-only circuits for nearest neighbor and next-nearest neighbor Pauli measurements. 
We separate the circuits by symmetry into orientable and non-orientable ones, and we validate their anticipated behavior based on symmetry.
In Sec.~\ref{ss:two_leg_ladder} (Sec.~\ref{sec:orient_non_gauss}), we explore Gaussian (non-Gaussian) orientable circuits and argue that their behavior can be understood in the framework of coupled 2D Potts models.
In Sec.~\ref{sec:non_orient_Gauss} (Sec.~\ref{sec:non_orient_non_gauss}) we in turn explore Gaussian (non-Gaussian) non-orientable circuits, and we show that in both cases they realize a CPLC Goldstone phase with universal properties.
In Sec.~\ref{sec:duality} we discuss spacetime duality for Majorana circuits and we argue that the measurement-only phases discussed in this work do not have a purely unitary spacetime dual counterpart. 
Finally, we consider the measurement-only XZZX quantum code in Sec.~\ref{ss:three_leg_ladders} and argue, based on geometry, that it is generally non-orientable and that it realizes a CPLC Goldstone phase.

\section{From 1D qubit circuits to 2D Loop Models}\label{sec:loop_review}

We consider measurement-only circuits of $L$ qubits, which are arranged in a one-dimensional chain. 
By virtue of the Jordan-Wigner (JW) transformation, such circuits may be represented in terms of Majorana fermions, including in the presence of parity-violating measurements.
We will argue, and subsequently numerically confirm, that this further enables a mapping of the circuit to the statistical mechanics of completely packed (dense) loop models in 2D~\cite{Fava_2023, Saleur_1986, Martins_1998, Jacobsen_2003, Candu_2009, Cardy_2002_annulus}.
For Gaussian Majorana operations, i.e., quadratic in Majorana fermions, this mapping is exact on the level of the microscopic circuit. 
For non-Gaussian circuits, the loop model is emergent and may depend on the complete set operators that is measured and on the parameter regime. 

In this section, we discuss the mapping of Majorana circuits to loop models based on (i) the Temperley-Lieb (TL) and Brauer algebras~\cite{Candu_2009_brauer} and (ii) the worldlines for the Majoranas.
We will then discuss a central aspect of these loop models: the orientability of Majorana worldlines. 
The presence or absence of orientability leads to fundamentally different long wavelength physics realized in the circuit.
We thus separate measurement-only circuits in the following discussion into orientable and non-orientable.
Orientable models can be studied by a number of well-established techniques ranging from the TL representation of the transfer matrix to Coulomb gas approaches.
In contrast, non-orientable loop models generically yield non-planar worldline configurations, termed completely-packed loop model with crossings (CPLC), which cannot be expressed by an element of the TL algebra. 
As one major consequence, they allow for a non-unitary symmetry broken phase in two dimensions, the \emph{Goldstone phase} of the CPLC~\cite{Nahum_2013, Jacobsen_2003}. 

We proceed in this section by including non-Gaussian measurements. 
We focus on two aspects: the relevance  in the renormalization group sense and the impact on the orientability of the loop model representation.

\begin{figure}[!ht]
    \centering
    \includegraphics[width=\columnwidth]{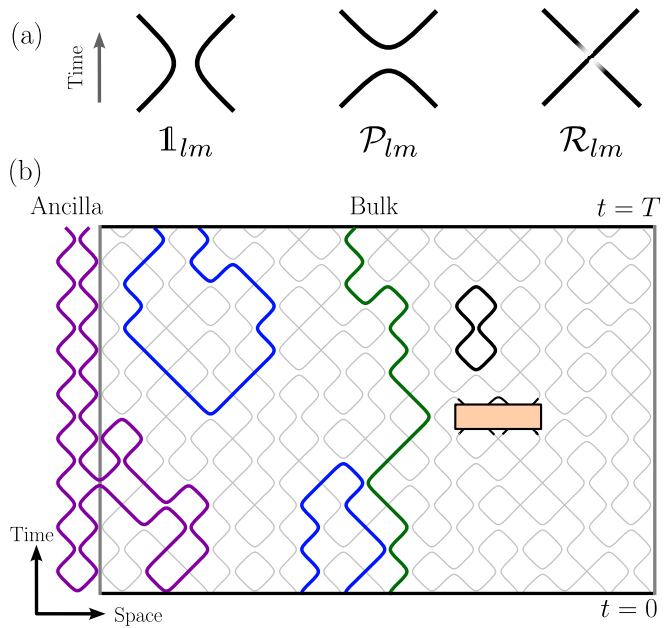}
    \caption{
    \textbf{Loop representation of Gaussian Majorana circuits: }
    (a)
    The three elementary operations on a pair of Majorana worldlines: The identity $\mathds{1}_{lm}$ propagates  unaffected loops in the temporal direction.
    A projective parity measurement $\mathcal{P}_{lm}$ connects loops in the opposite way, appearing like spatial propagation. A swap gate $\calR_{lm}$ makes the worldlines cross over one another.
    (b)
    A loop configuration for brickwall Majorana circuit of depth $T$. Different types of loops contribute to different correlation functions:
    Spanning loops (green) connect the two temporal boundaries, fixing boundary correlation functions like $\langle \gamma_i(t=0) \gamma_j (t=T) \rangle$.
    Loops with both ends terminating on the same temporal boundary (blue) determine the equal-time spatial correlations $\langle \gamma_i(t)\gamma_j(t)\rangle\vert_{t=0,T}$.
    Adding ancilla Majorana modes allows loops (purple) to cross the spatial boundaries.
    Closed loops (black) contained entirely in the spacetime bulk do not contribute to boundary observables but modify bulk correlation functions.
    Non-Gaussian, e.g., four-fermion, measurements are represented as multi-leg vertices (orange box) which are not immediately expressed as a single loop configuration.
    }
    \label{fig:loop_overview}
\end{figure}

\subsection{Mapping Discrete Gaussian Majorana Circuits to Loop Models}\label{sec:loop_circuits}
Here we introduce the mapping between the statistical mechanics of two-dimensional loops and the dynamics of $(1+1)$-dimensional, discrete Gaussian Majorana circuits.
We label the $L$ qubits in the circuit by an index $j=1,...,L$.
Each qubit $j$ admits an alternative representation in terms of two Majorana fermions $\gamma_{2j-1}, \gamma_{2j}$, which fulfill the anti-commutation relation $\{\gamma_{l}, \gamma_{m}\}=2\delta_{l,m}$ for $l,m=1,...,2L$. 
The circuit evolves under measurements of Pauli strings $\hat O$, i.e., each operator $\hat O$ is a string $\hat O=\otimes_{j=1}^L\hat A_j$ with $\hat A_j\in \{\mathds{1}_j, X_j, Y_j, Z_j\}$ being a Pauli operator or the identity acting on qubit $j$. 

Under the JW transformation, each Pauli string yields a corresponding Majorana operator which falls into one of three categories: 
(i) \emph{Gaussian} operators $\hat O=i\gamma_l\gamma_m$ consisting of exactly two Majorana modes,
(ii) parity preserving but non-Gaussian operators $\hat O= i^n \prod_{l=1}^{n>1} \gamma_{j_{2l-1}}\gamma_{j_{2l}}$ consisting of an even number of Majorana modes and
(iii) parity breaking operators $\hat O =\gamma_{j_{2l+1}}i^n \prod_{l=1}^{n\ge0} \gamma_{j_{2l-1}}\gamma_{j_{2l}}$ consisting of an odd number of Majorana modes. 

We start with the Gaussian case, i.e., $\hat O=i\gamma_l\gamma_m$, and introduce the projector $\calP_{lm}\equiv\tfrac12(\mathds{1}+\hat O)=\tfrac12(\mathds{1}+i\gamma_l\gamma_m)$ onto the positive measurement outcome $\hat O=+1$. 
The opposite outcome is obtained by exchanging $l\leftrightarrow m$. 
Similarly, we introduce the unitary  $\calR_{lm}\equiv \exp\left(i\tfrac{\pi}{4}\hat O\right)=\tfrac{1}{\sqrt{2}}(\mathds{1}-\gamma_l\gamma_m)$, which swaps Majoranas $\gamma_l\leftrightarrow\gamma_m$.
Upon normalization of the state after each projection, for four different Majorana fermions $\gamma_n, \gamma_s, \gamma_l, \gamma_m$ this yields the algebra: 
\begin{align}
    0&=[\calP_{lm},\calP_{ns}]=[\calR_{lm},\calR_{ns}],\label{eq:commutation}\\
    \calP_{lm}&=\calP_{lm}^2=\calP_{lm}\calP_{nl}\calP_{lm}, \label{eq:projector_algebra}\\
     \calP_{lm}&=\calR_{ms}^\dagger\calP_{ls}\calR_{ms}, \ \calR_{lm}=\calR_{ms}^\dagger\calR_{ls}\calR_{ms}, \label{eq:swap-proj-algebra} \\
     \calR_{lm}^4 &= \mathds{1}. \label{eq:swap_powers}
\end{align}
We emphasize that the last equality in Eq.~\eqref{eq:projector_algebra} is not an operator identity (which would require a factor $\frac12$) but holds when the quantum state on which the projectors are acting is normalized after each measurement. 

The algebra is general and holds for any combination of Majorana indices. 
If we restrict the projectors $\calP_{lm}$ to nearest neighbor pairs $|l-m|=1$ and exclude swaps, then Eqs.~\eqref{eq:commutation}-\eqref{eq:projector_algebra} yield the paradigmatic Temperley-Lieb (TL) algebra.
The TL algebra describes families of exactly solvable loop models, here with loop fugacity $n=1$, such as the $1$-state Potts model~\cite{Martin_1991_Potts, Wu_1982_review}. 
If we include nearest neighbor swaps $\calR_{lm}$ with $|l-m|=1$, Eqs.~\eqref{eq:commutation}-\eqref{eq:swap-proj-algebra} reveal yet another loop model structure, known as the Birman-Murakami-Wenzl (BMW) algebra~\footnote{In particular, loops with fugacity $n=1$ can be described by the $SO(4)_1$ representation of the BMW algebra.}\cite{Fendley_2008, Birman_1989, Murakami_1987}.

Importantly, lifting the constraint $|l-m|=1$ and allowing measurements and swaps $\calP_{lm}, \calR_{lm}$ of arbitrary Majorana pairs $\gamma_l,\gamma_m$, all operations remain well-described by the BMW algebra.
By virtue of Eq.~\eqref{eq:swap-proj-algebra}, any non-local swap $\calR_{lm}$ or measurement $\calP_{lm}$ with $|l-m|>1$ can be expressed via a sequence of nearest-neighbor swaps, e.g., $\calP_{1,3}=\calR_{2,3}^\dagger\calP_{1,2}\calR_{2,3}$.
The BMW algebra thus provides a framework for arbitrary measurement-only circuits of Gaussian Majorana operators $\hat O=i\gamma_l\gamma_m$.

Dense loop models naturally correspond to diagrammatic representations of the TL or BMW algebra, which for Gaussian Majorana circuits are quite intuitive.
Each Majorana fermion is represented by a worldline on a (1+1)-dimensional lattice with a spatial and a temporal coordinate, as depicted in Fig.~\ref{fig:loop_overview}. 
For each timestep $t$ and neighboring pair $l,m$ on the tilted square (brickwall) lattice, one of the following generators of the BMW algebra is applied, as shown in Fig.~\ref{fig:loop_overview}a.
The identity $\mathds{1}_{lm}$ propagates Majorana fermions $\gamma_l, \gamma_m$ unmodified in time.
A parity measurement $\calP_{lm}$ of the pair $\gamma_l, \gamma_m$ fixes the parity $i\gamma_l\gamma_m$ and connects the worldlines $l,m$ to form an avoided crossing in time.
A swap $\calR_{lm}$  exchanges the Majorana fermions $\gamma_l \leftrightarrow \gamma_m$, and corresponds to a non-avoided crossing in space-time. 

A selection of possible space-time loops, including potential boundary conditions, is shown Fig.~\ref{fig:loop_overview}b.
As one infers from the diagrams, the loop representation makes no distinction between $\calP_{lm}$ and $\calP_{ml}$, i.e., it is agnostic to the sign of each parity measurement outcome. 
Thus, each loop diagram represents an equivalence class of circuits rather than a specific circuit realization. 
Each circuit of a given equivalence class is identical up to the random measurement outcomes and the sign of the swap gates. 
Consequently, each equivalence class predicts quantities that are independent of measurement outcomes. 
Examples include the modulus $\abs{i\langle\gamma_l\gamma_m\rangle}=0,1$ of each parity, the entanglement entropy and the purity of the state.

This equivalence relation identifies both signs of swap gates, $\calR_{lm} \sim \calR_{lm}^\dagger$ such that Eq.~\eqref{eq:swap_powers} is replaced by $\calR_{lm}^2 \sim \mathds{1}$.
Then the relation between the operations reduces to the Brauer algebra, i.e., the degenerate point of the BMW algebra.
The Brauer algebra $\calB_{N}$ consists of pairwise matchings between $2N$ elements, which we can identify with the endpoints of Majorana worldlines at temporal boundaries of the circuit.
Then the elements of $\calB_N$ are in one-to-one correspondence with equivalence classes of Gaussian Majorana circuits, specified by the pairings $\abs{i\langle \gamma_l \gamma_m \rangle}$.
Two circuits from the same equivalence class can be transformed into one another by applying a series of local phase gates, which do not change the entanglement. 
This is true for general Clifford circuits~\cite{Fattal_2004} and enables an experimental procedure to detect the predicted observables~\cite{Noel_2022}. 

\begin{table}[t]
    {
    \centering
    \setlength{\tabcolsep}{1.5em}
    \setlength{\extrarowheight}{0.25em}
    \begin{tabular}{|c|c|}
        \hline
        \textbf{State} & $\ell^2 {\mathbf{P(\ell)}} \left[\frac{1}{3\log(2)}{\mathbf{\tilde{c}}}(\ell)\right]$ \\ [0.5ex]
        \hline \hline
        Critical $Q$-state Potts~\cite{Jacobsen_2008} & $\frac{(2 + \sqrt{Q})}{\pi\sqrt{4-Q}} \frac{\arcsec(2/\sqrt{Q})}{\arcsec(-2/\sqrt{Q})}$ \\
         CPLC critical point~\cite{Nahum_2013} & $\frac{2.035}{\pi}$ \\
         CPLC Goldstone phase~\cite{Nahum_2013} & $\frac{1}{2\pi^2}\log(l) + \textrm{const.}$ \\
         \hline
    \end{tabular}
    }
    \caption{Asymptotic behavior of the loop-length distribution for the long-loop phases of CPLC and the critical point of the Q-state Potts models. It determines the prefactor of the von Neumann entanglement entropy $S_{L/2}=\tfrac{\tilde c(L)}{3}\log(L)$.}
    \label{tab:ref_table}
\end{table}

\subsection{Circuit Observables and Loop Configurations}\label{sec:observables}

The dynamics of a circuit are characterized by bulk and boundary observables. 
The latter are correlations at the temporal boundary, i.e., in the final state $\rho(T)$, such as the entanglement entropy.
When the circuit has a rigorous loop model representation, $\rho(T)$ can be inferred from boundary loop configurations. 
Then it is more efficient to simulate the corresponding loop model rather than the Clifford circuit (see App.~\ref{app:computation_details} for details). 
When no such microscopic mapping to a loop model is possible, we instead simulate the Clifford circuit via the tableau formalism~\cite{Aaronson_2004}. 
Below we provide a dictionary connecting observables in Clifford circuits and in loop models. 

In a loop model, boundary observable are encoded in the distribution of fermion parities $\abs{\langle\gamma_m\gamma_l\rangle_T}$ at time $T$.
Each non-zero parity corresponds to a loop arc connecting sites $l,m$ with the length of the arc defined as $\ell=|l-m|$~\footnote{This should not be confused for the total length of the loop through the whole spacetime bulk, as considered in Ref.~\onlinecite{Nahum_2013}.}. 
The entanglement properties and correlations in $\rho(T)$ are captured by the normalized loop length distribution $P(\ell)$, which is in one-to-one correspondence with the stabilizer length distribution in the clipped gauge~\cite{Nahum_2017, Li_2019}.
At a critical point of the loop model, the loop length distribution takes the form $P(\ell)=\tilde c(\ell) \tfrac{3\log(2)}{\ell^2}$ with a universal prefactor $\tilde c(\ell)$. 
In Table~\ref{tab:ref_table} we show the known values of $\tilde{c}(\ell)$ for both CPLC and the $Q$-state Potts model.

\emph{Entanglement Entropy and Mutual Information. } --
From the state $\rho(T)$ we compute the von Neumann entanglement entropy $S_A \equiv -\Tr\left[\rho_A(T) \log_2 \rho_A(T) \right]$ of a subregion $A$ and the mutual information $I_2(A,B) \equiv S_A + S_B - S_{AB}$ between two subregions $A,B$. 
Here $\rho_A(T)=\Tr_{\bar A}(\rho(T))$ is the reduced density matrix obtained by tracing out the complement $\bar A$ of $A$. 
In a Gaussian circuit, both $S_A$ and $I_2(A,B)$ are determined by the set of fermion parities $|\langle \gamma_l\gamma_m\rangle_T|=1$, i.e., by the loop arcs at $t=T$. 
Then $S_A=N_A$~\footnote{Please note that we work in units of $\log(2)$ so that entropies reflect exactly the integer number of loops.} is the number of arcs $N_A$ that terminate with one end in $A$ and with the other end in its complement $\bar A$~\cite{Sang_2021, Lang_2020, Fattal_2004}. 
For a loop distribution $P(\ell)$ with scaling $\ell^2 P(\ell)=3\log(2)\tilde{c}(\ell)$ as in Table~\ref{tab:ref_table}, this is $N_A=\tilde{c}(|A|) / 3\log(2)$ (see App.~\ref{app:computation_details}).
The mutual information $I_2(A,B)$ is twice the number of loops with one endpoint in $A$ and the other endpoint in $B$.
For Gaussian circuits and loop models, mutual information is additive, i.e., $I_2(A, BC) = I_2(A,B) + I_2(A,C) - I_2(A, B \cap C)$~\footnote{The mutual information $I_2(A,B)$ is uniquely characterized by the ensemble average of $\sum_{l\in A}\sum_{m\in B}|\langle \gamma_l(T) \gamma_m(T) \rangle|$.}. 

In order to detect long-range entangled states in the circuit, we compute the mutual information between contiguous subsystems $A,B$ of size $\abs{A} = \abs{B} = N/8$ Majorana fermions with separation $N/8$ (e.g. $A = [1, N/8]$ and $B = [N/4 + 1, 3N/8]$). 
For this choice, $I_2(A, B)$  provides a convenient metric for the finite-size scaling analysis in the vicinity of a critical point. 
It can be used for both locating an entanglement transition and determining the critical exponent $\nu$~\cite{Lavasani_2021, Klocke_2022}.

\emph{Multipartite entanglement.} -- 
For a Gaussian circuit, i.e., for loop models, every open loop has precisely two endpoints~\cite{Sang_2021} and zero operator content between them. 
Thus any multipartite entanglement can be expressed by the sum of bipartite mutual information. 
For instance the tripartite mutual information $I_3(A,B,C)$ always vanishes in a loop model. 
This is different for generic Clifford circuits: while for contiguous and adjacent subsystems $A$ and $B$, the entanglement entropy of a stabilizer state is the same as for the corresponding Majorana wave function, this is no longer true when computing the mutual information between \emph{disconnected} regions
\footnote{
As a minimal example, consider the four qubit stabilizer state with stabilizer generator $\mathcal{G} = \langle \gamma_1\gamma_5, \gamma_2\gamma_3, \gamma_4\gamma_8, \gamma_6\gamma_7\rangle$. After a Jordan-Wigner transformation we have $\mathcal{G} = \langle X_1 X_2, Y_1 Z_2 X_3, X_3 X_4, X_2 Z_3 Y_4 \rangle$ in the clipped gauge. Consider now taking subsystems $A,B,C$ to be qubits 1, 2, and 4, respectively. Whereas $I_2(B,C) = 1$ from the fermionic loop representation, in the spin-qubit language we have $I_2(B,C) = 0$.
As a result, the tripartite mutual information $I_3(A,B,C)$ is non-vanishing in the spin language, while it is necessarily zero in the fermionic loop language.
The origin of this difference lies in the non-local nature of the Jordan-Wigner transformation.
Stabilizers with endpoints in disconnected regions such as $B$ and $C$ still have non-trivial operator weight in the intervening region due to the Jordan-Wigner string.
}.

\emph{Purity and erasure of initial information.} -- 
For a circuit with Majorana worldline representation, the \emph{spanning number} $n_s(T)$ counts the number of worldlines connecting the $t=0$ and $t=T$ temporal boundaries of the circuit. 
It is the mutual information between the initial and the final state and thus quantifies the survival of information inserted into the circuit at $t=0$ up to the final time $t=T$.
We distinguish two important cases: 
Firstly, when the initial state is maximally mixed, it hosts only unpaired Majorana worldlines and the spanning number counts all unpaired Majorana world lines at time $T$. 
This provides the total entropy $S_L(T) = -\Tr\left[\rho(T) \log_2 \rho(T) \right] = \tfrac12 n_s(T)$, i.e., the time-dependent purity.
Secondly, when the initial state is pure, $n_s(T)$ is the number of Majorana parities which have not been erased by measurements. 
Then it measures the survival probability of initially encoded information. 
This elucidates the equivalence between dynamical purification of an initial mixed state and the erasure of information from an initial pure state by measurement.

\emph{Bulk observables.} -- 
Both 2D loop models and (1+1)D circuits hold information that is not accessible from boundary correlations. 
In loop models, an important example is the so-called watermelon correlator $G_{2n}(\vec{z}, \vec{z}')$. 
For any integer $n$ it is the probability that the (two-dimensional) coordinates $\vec{z}, \vec{z}'$ in the bulk of the loop configuration are connected by $n$ \emph{closed loops}.
In the circuit $\vec{z}=(l,t), \vec{z}'=(l', t')$ become space-time coordinates and, e.g., $G_{2}(\vec{z}, \vec{z}')$ is the probability that the two Majorana fermions $\gamma_l, \gamma_{l'}$ share the same worldline for intermediate times $0 < t,t' < T$ and $t\neq t'$.
We express this as the parity $|\langle \gamma_{l}(t)\gamma_{l'}(t')\rangle|=1$, which takes the form of an out-of-time order correlator. 
It reveals information about the dynamics that boundary correlations cannot access.
In Appendix~\ref{sec:ancilla_probe} we show how this bulk quantity can be extracted from Gaussian and general Clifford circuits.

\subsection{Worldline orientability, bipartite lattices and time-reversal symmetry}\label{sec:orientability}

\begin{figure}[t]
    \centering
    \includegraphics[width=\columnwidth]{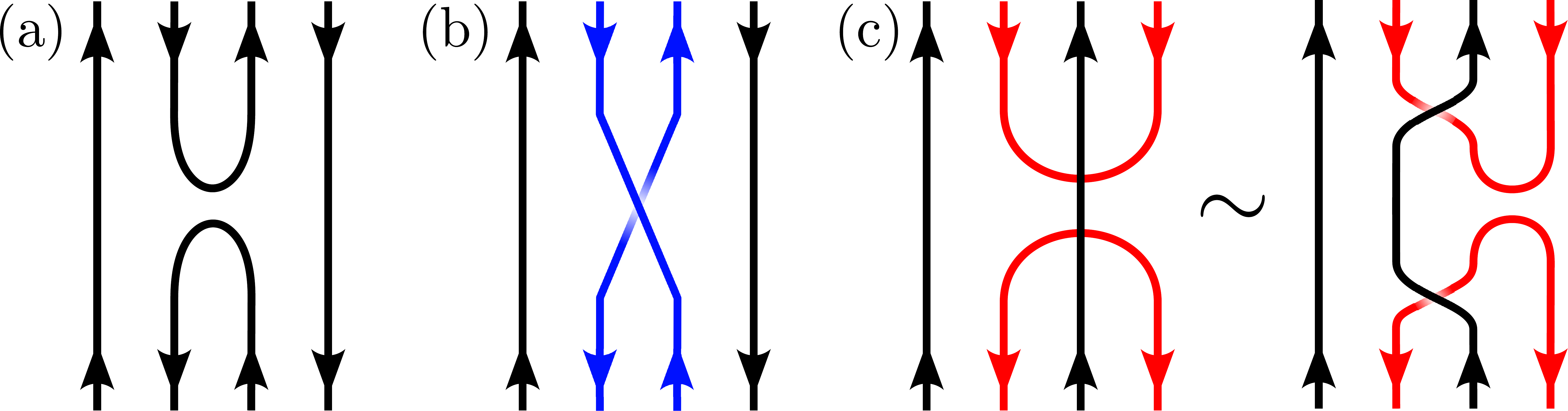}
    \caption{
    \textbf{Orientability of worldlines.}
    Worldlines get assigned a fixed orientation for the even or the odd sublattice. (a) Nearest neighbor measurements preserve this orientation.
    By contrast, (b) nearest-neighbor crossings and (c) next-nearest neighbor measurements mix the orientations and break the associated symmetry.
    }
    \label{fig:worldline_orientations}
\end{figure}

A core aspect of the statistical mechanics of loop models and of measurement-only circuits is the so-called orientability of the loops.
In order to introduce the concept of orientability, let us start with a TL algebra resulting from only nearest neighbor measurements $\calP_{l,l+1}$.
We then let all odd Majorana worldlines $\gamma_{2l-1}$ furnish a representation of $U(1)$ while all odd worldlines $\gamma_{2l}$ furnish the adjoint representation, corresponding to forward and backward evolution in time~\cite{Candu_2009}.
Then the TL generators $\calP_{l,l+1}$ are invariant under action of a $U(1)$ symmetry~\footnote{A way to define a symmetry operator for orientability is provided in Ref.~\cite{Bulgakova_2020}. Consider the TL algebra with a bipartition between odd and even Majoranas. Define $Q \equiv \sum_i \left( \calP_{2i-1, 2L} + \calP_{2i, 2L-1} - \calR_{2i-1,2L-1} - \calR_{2i, 2L}\right)$. This commutes with all $\calP_{2i-1, 2j}$ but does not commute with $\calP_{2i, 2j}$ or $\calP_{2i-1, 2j-1}$.}.
As shown in Fig.~\ref{fig:worldline_orientations}a, the application of $\calP_{l,l+1}$, preserves the local orientations.

Extending the TL to the Brauer algebra by including swaps $\calR_{l,l+1}$ changes the situation: nearest neighbor swaps exchange even and odd worldlines and thus locally violate orientability, see Fig.~\ref{fig:worldline_orientations}b. 
By contrast, a next-nearest neighbor swap $\calR_{l, l+2}$ respects the alternating orientations and leaves the $U(1)$ symmetry intact~\cite{Candu_2009}.
Since next-nearest neighbor measurements $\calP_{l,l+2}$ can be expressed as a measurement $\calP_{l+1,l+2}$ conjugated by a swap $\calR_{l, l+1}$, they also violate orientability, as shown in Fig.~\ref{fig:worldline_orientations}c.
More generally, any measurement $\calP_{l,l+m}$ with $m$ odd (even) preserves (violates) the orientation, whereas any swap $\calR_{l,l+m}$ with $m$ even (odd) preserves (violates) the orientation.

This construction identifies orientability as the counterpart of time-reversal symmetry in Majorana circuits. 
To see this, we first realize that a projection $\calP_{l,l+m} = \tfrac12(\mathds{1} + i\gamma_l\gamma_{l+m})$ with $m$ odd will always connect a Majorana with an even index to a Majorana with an odd index. 
By convention, odd Majoranas are real $\gamma_{2l-1}\sim c^\dagger_l+c_l$, while even Majoranas are complex $\gamma_{2l}\sim i(c^\dagger_l-c_l)$.
Thus products $i\gamma_l\gamma_{l+m}$ with $m$ odd are real and invariant under time reversal.
The same is true for swaps $\calR_{l,l+m}=\tfrac{1}{\sqrt{2}}(\mathds{1}-\gamma_l\gamma_{l+m})$, where for $m$ even the product $\gamma_l\gamma_{l+m}$ is real. 
When all the generators of the evolution are invariant under time-reversal, so is the entire circuit.
The symmetry implies that all circuit observables that are odd under time-reversal must have zero expectation value.

This framework can be further generalized: a given circuit has a time-reversal invariant representation and it preserves orientability whenever we can find a bipartition into two sublattices $A, B$ such that 
(i) all measurements are of the form $\calP \sim \mathds{1} + i \gamma_{A_l} \gamma_{B_m}$ with $A_l$ ($B_m$) being on sublattice $A$ ($B$), i.e., measure \emph{inter-sublattice parities} and 
(ii) all swaps are of the form $\calR \sim \mathds{1} -\gamma_{A_l} \gamma_{A_m}$ or $\calR \sim \mathds{1} - \gamma_{B_l} \gamma_{B_m}$, i.e., exchange Majoranas only within a given sublattice.
The proof is straightforward: assign all worldlines on sublattice $A$ a positive orientation and all worldlines on $B$ a negative orientation. 
This provides a practical criterion for orientability in a loop model. 
If all (any) closed loops, see black configuration in Fig.~\ref{fig:loop_overview}, originate from an even (odd) number of measurements, the circuit is bipartite (non-bipartite).
This provides the equivalence of (i) the bipartiteness of a Majorana circuit, (ii) time-reversal invariance of the generators $\calP, \calR$ and (iii) orientability of the corresponding loop model.

For an orientable loop model with $N$ worldlines, and with a bipartition $A,B$, the difference between positively and negatively oriented worldlines $Q=|A|-|B|\in \mathds{Z}$ with $-N\le Q\le N$ is the conserved charge under $U(1)$~\footnote{More precisely, globally conserved operators can be defined by taking elements of the center of the group algebra $\calB_{\abs{A}, \abs{B}}.$}.
The loop configurations are elements of the \emph{walled} Brauer algebra $\calB_{|A|, |B|} \subset \calB_{N}$~\cite{Bulgakova_2020, Benkart_1994_walled_brauer}.

\subsection{Effective Long-Wavelength Description}

Unambiguously mapping a monitored circuit to a loop model gives rise to a large toolbox of theoretical methods to examine its long wavelength behavior. 
An important example is the mapping of dense loops to nonlinear sigma models (NL$\sigma$Ms)~\cite{Jacobsen_2003, Nahum_2013, Fava_2023}, whose form and universal behavior depend explicitly on the presence or absence of orientability. 
When the local orientations are conserved, the order parameter takes values in $\mathbb{CP}^{n-1}$. 
If instead orientability is broken, as in CPLC, the order parameter is reduced to $\mathbb{RP}^{n-1}$~\cite{Nahum_2013}. 
The relevant case of $n=1$ may be studied as a replica-like limit of the $n > 1$ theory~\footnote{Or alternatively by a supersymmetric formulation on an enlarged space $\mathbb{CP}^{m|m}$ with $m \in \mathbb{Z}_+$\cite{Candu_2009}}.

Once orientability is broken sufficiently strongly, the corresponding NL$\sigma$M features a phase transition from a short-loop phase to a symmetry-broken, so-called Goldstone phase. 
It is an order-disorder transition driven by $\mathbb{Z}_2$ vortex proliferation in the $\mathbb{RP}^{n-1}$ theory~\cite{Jacobsen_2003, Nahum_2013, Fava_2023}.
Despite taking place in two-dimensions, this transition and the CPLC Goldstone phase are not prohibited by the Mermin-Wagner theorem since a non-hermitian representation is required for treating the $n=1$ loop model~\cite{Jacobsen_2003, Nahum_2013}. 
Whenever orientability is broken in the circuit, we will resort to the CPLC NL$\sigma$M results, which are universally confirmed in this case. 

\subsection{Symmetry Preserving Non-Gaussian Perturbations}\label{sec:beyond_ff}

The mapping of Gaussian Majorana circuits to loop models was based on the TL and Brauer algebra in Eqs.~\eqref{eq:commutation}-\eqref{eq:swap-proj-algebra}, which is rigorous for Gaussian circuits.
We show below that measuring non-Gaussian operators creates superpositions of loop configurations. 
This is analogous to Hamiltonian systems, where interactions create superpositions of Gaussian states. 
In many situations, however, a unique loop model may emerge on a coarse grained scale. 
Here we adopt a perspective inspired by the renormalization group (RG) framework and consider measurements of non-Gaussian operators $\{\hat O_l\}$ as perturbations on top of an otherwise Gaussian circuit. 
We note that this situation is quite general: the frustration graph of an arbitrary set of measured operators always contains a subgraph that can be represented by measurements of Gaussian fermion operators~\cite{Elman_2021}. 
The remaining operators then act as non-Gaussian perturbation.

\begin{figure}[t]
    \centering
    \includegraphics[width=\columnwidth]{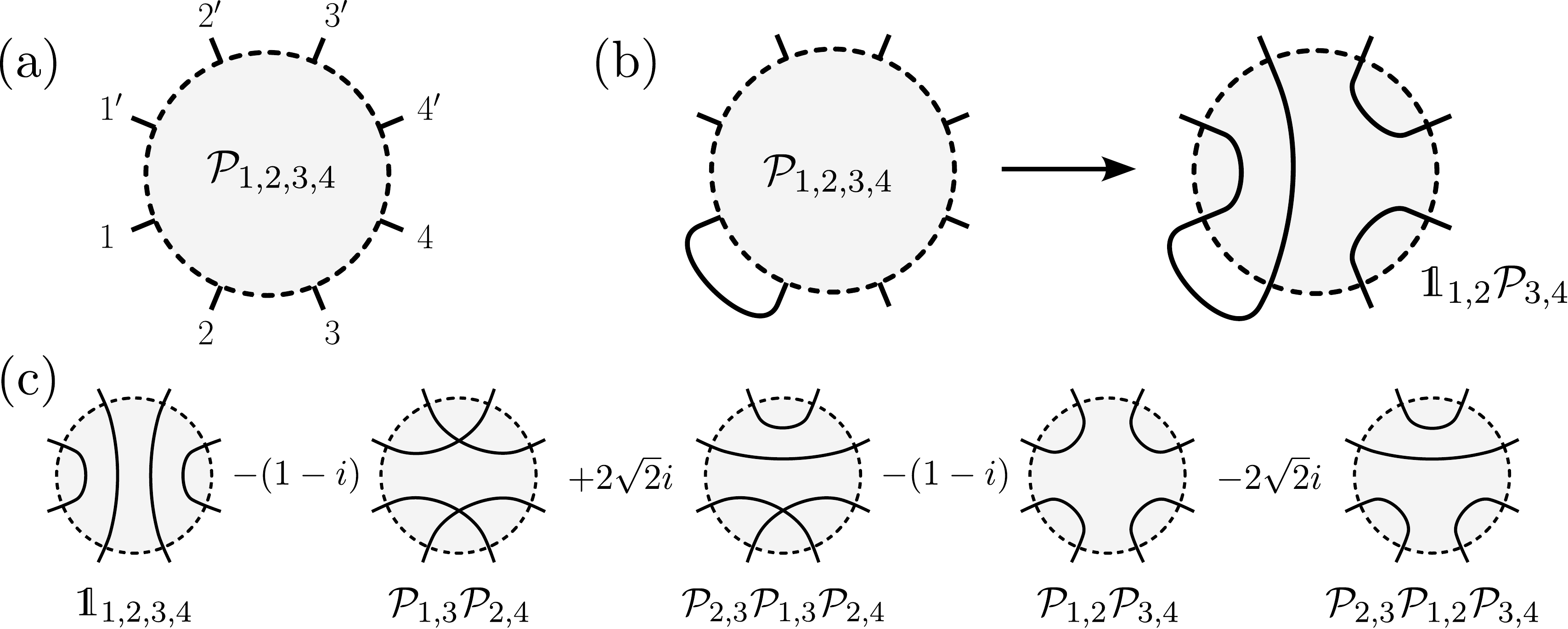}
    \caption{
    \textbf{Four-fermion measurements.} 
    (a) Measuring the total parity on four Majoranas $\gamma_{l}$ for $l=1,2,3,4$ can be represented by an 8-leg vertex, where the unprimed and primed indices denote incoming and outgoing worldlines, respectively.
    (b) When one of the local fermion parities is known due to a closed loop entering the vertex, the measurement reduces to a parity-check on two fermions, restoring a vertex-free loop diagram.
    (c) Without knowledge of the local parities, the vertex represents a sum over possible pairings of the legs.
    One of several equivalent diagrammatic representations is shown here.
    Notably, the sum includes configurations which are inconsistent with orienting the worldlines.
    }
    \label{fig:four_fermion_vertex}
\end{figure}

Starting from a Gaussian circuit, i.e., a well-defined 2D loop model, the notion of relevance and irrelevance in the RG sense for non-Gaussian perturbations in 2D statistical mechanics is applicable. 
Thus, measuring a set of non-Gaussian operators $\{\hat O_l\}$, which preserve the symmetries of the Gaussian circuit, is an irrelevant perturbation when added to an extended thermodynamic phase of either short or long loops.
It may, however, become relevant near a Gaussian critical point and modify the critical behavior of the circuit analogously to a Wilson-Fisher fixed point. 
Further, when the circuit becomes dominated by non-Gaussian measurements it may approach a strong coupling fixed point.

The effect of symmetry-preserving non-Gaussian measurements in a loop model can be illustrated by starting with a Gaussian circuit described by the Brauer algebra in Eqs.~\eqref{eq:commutation}-\eqref{eq:swap-proj-algebra}.
It shall be perturbed by measurements of non-Gaussian operators which are symmetry-preserving, e.g., preserve parity and orientability and are thus even under time-reversal.
Let us take for example four-fermion parity checks, such as $Z_l Z_{l+1}=\gamma_{2l-1} \gamma_{2l} \gamma_{2l+1} \gamma_{2l+2}$ or $ X_l X_{l+2} = \gamma_{2l} \gamma_{2l+1} \gamma_{2l+2} \gamma_{2l+3}$. 
The projective measurement $\calP_{1,2,3,4} = \tfrac12 (\mathds{1} - \hat{O}_{1,2,3,4})$ of a general four-fermion operator $\hat O_{1,2,3,4}=\gamma_1\gamma_2\gamma_3\gamma_4$ corresponds to an ``8-leg vertex'' in the loop model.
It has four incoming and four outgoing worldlines (see Fig.~\ref{fig:four_fermion_vertex}a), which, in contrast to two-fermion measurements, have \emph{no} unambiguous connection between individual pairs of fermions.
Instead, $ \calP_{1,2,3,4}$ yields a superposition of all possible ways of connecting the legs incident on the vertex with a fixed total parity. 
This gives rise to several, equivalent representations of $\calP_{1,2,3,4}$ in the loop framework. 
For instance, the projector can be written as
\begin{align}\label{eq:superposition}
        \calP_{1,2,3,4} = \mathds{1} &-i\sqrt{2}\left(e^{-\frac{i\pi}{4}}\mathds{1} - \sqrt{2}\calP_{2,3}\right)\calP_{1,3}\calP_{2,4} \\ 
        &+ i\sqrt{2}\left(e^{+\frac{i\pi}{4}}\mathds{1} - \sqrt{2}\calP_{2,3}\right)\calP_{1,2}\calP_{3,4}.\nonumber
\end{align}
The corresponding superposition of loop configurations is shown in Fig.~\ref{fig:four_fermion_vertex}c. 
So far, the sum is \emph{not} restricted to terms which respect the symmetry, e.g., orientability, of the Gaussian circuit and so may include both symmetry preserving and symmetry breaking terms.

We may now ask what the action of Eq.~\eqref{eq:superposition} will be in an otherwise Gaussian circuit.
Whenever the parity of two of the four fermions is already known \emph{or} measured in the future, the four-fermion measurement reduces to a two-fermion parity check, as revealed from  $\calP_{1,2}\calP_{1,2,3,4}=\calP_{1,2,3,4}\calP_{1,2}=\calP_{1,2}\calP_{3,4}$ and depicted in Fig.~\ref{fig:four_fermion_vertex}b. 
This provides access to two important limits: 
(i) When the circuit is dominated by local Gaussian measurements, an extensive number of local parities is known immediately before or after application of $\calP_{1,2,3,4}$. 
Then $\calP_{1,2,3,4}$ can almost always be replaced by two-fermion measurements.
(ii) In turn, when non-Gaussian measurements are dominant, few Gaussian parity checks will induce a cascade of fixed local parities, which collapses the circuit onto a short loop state.
Both cases give rise to a loop model interpretation, in which the topology of the short loop (area law) phase depends on (i) the Gaussian or (ii) the non-Gaussian measurements. 

This observation is readily generalized to measurements of arbitrary symmetry-preserving Majorana operators $\hat O\sim \gamma_1,...\gamma_{2n}$ with $n>2$. 
When the parity of $m$ bilinears in $\hat O$ is known, then it reduces to the measurement of an $2(n-m)$ Majorana operator.
The size of the reduced operator corresponds precisely to the mutual information between the support of $\hat{O}$ and its complement in the circuit.
This mutual information is a reliable indicator of the relevance of a $2n$-fermion operator in a Gaussian circuit. 
In an area law phase with short loops, and for contiguous support of $\hat O$, it will not grow with $n$. 
At a critical point with long loops (logarithmic entanglement growth), it will grow $\sim \log(n)$, such that higher-order operators become increasingly relevant. 
In this case, the $2n$-fermion measurements will not collapse onto a single loop configuration but instead will yield a superposition of all symmetry-allowed configurations.

A further important consequence of the above discussion is that symmetry-preserving non-Gaussian measurements indeed inherit the symmetry of the Gaussian circuit. 
Consider for instance orientability: when the worldlines entering the $2n$-fermion vertex form an orientable, bipartite lattice, the corresponding vertex is collapsed onto its bipartite part and cannot break orientability. 
In turn, when the worldlines instead form a non-bipartite lattice, the $2n$-fermion vertex will collapse onto its non-orientable part. 
Thus $2n$-fermion parity checks, which are compatible with orientability, i.e., even under time-reversal, will inherit the (non-) orientable structure of the underlying Gaussian circuit.

\subsection{Parity Violating Measurements}\label{ss:parity_breaking}

A second important class of non-Gaussian operations appearing in generic Clifford circuits are those which violate the total parity $\bar{Z} = \prod_{l=1}^L Z_l$. 
In general, a parity breaking Pauli operator, e.g., a single $X$ or $Y$, maps to an odd-length string of Majorana fermions attached to one (w.l.o.g the left) spatial boundary of the circuit. For Clifford circuits, two consecutive parity violating measurements, i.e., $X_l, X_m$ can be expressed as a parity conserving, $X_lX_m$, and a parity violating, $X_m$, measurement. 
We show that in the Majorana framework, this process is equivalent to an effective long-range swap $\calR_{2l, 2m-1}$.
Depending on the measured operators, their spatial position, and the time between measurements, such swaps are non-local in space-time and may bridge large distances.
Moreover, these swaps yield a potential mechanism for violating orientability, allowing a CPLC Goldstone phase to emerge.

\begin{figure}[t]
    \centering
    \includegraphics[width=\columnwidth]{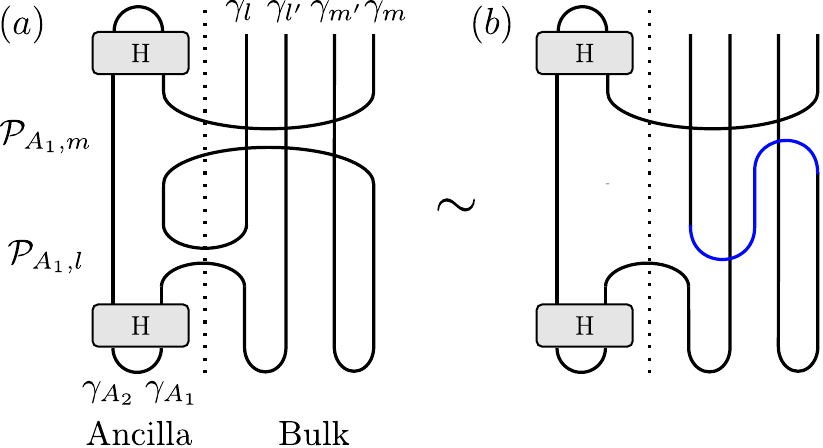}
    \caption{
        \textbf{Ancilla-mediated parity non-conserving measurements:}
        (a)
        Two sequential single Majorana measurements mediated via an ancilla qubit, $\calP_{A_1, l}$ at time $t_1$ and $\calP_{A_1, m}$ at time $t_2$.
        At the end of the evolution, the ancilla is decoupled from the bulk by a measurement in the $X$ basis.
        (b)
        An equivalent circuit generating the same final state involves a connection (blue) of bulk loops that is non-local in both space and time.
        This can be viewed as an effective swap $\calR_{l,m}$ after the first measurement $\calP_{A_1, l}$.
        If $\gamma_l$ and $\gamma_{m}$ are in separate bipartitions of the Majoranas, this effective swap facilitates a breaking of the worldline orientability.
    }
    \label{fig:ancilla_measurement}
\end{figure}

To include parity breaking measurements in the Majorana framework, we recast them as parity-preserving ones by introducing two ancilla fermions~\cite{Bravyi_2002}.
We place an ancilla qubit $A$ consisting of two Majorana modes $\gamma_{A_1}, \gamma_{A_2}$ at the left boundary of the system, as in Fig.~\ref{fig:ancilla_measurement}.
Performing a $Z_A$-measurement followed by a Hadamard gate, the ancilla is prepared in an eigenstate $\langle X_A \rangle = 1$ without well-defined parity. 
For a parity non-conserving operator $\hat{O}$ acting in the bulk, prepending $X_{A}$ yields a new operator $\hat{O} \rightarrow X_{A} \hat{O} = \gamma_{A_1} \hat{O}$ that is even in Majorana modes. 
This now conserves the parity of the joint system, $Z_A \bar{Z}$ and leaves the bulk unmodified compared to measuring only $\hat O$.

To illustrate the effect of the ancilla, we consider the following simple example. 
Consecutively measure two single Majorana fermions: $\hat O=\gamma_l$ at time $t$ and $\hat O'=\gamma_{l'}$ at time $t'$. 
In the ancilla framework, this yields the Gaussian projections $\mathcal{P}_{A_1, l} = \tfrac12(1 + i\gamma_{A_1}\gamma_l)$ at time $t$ and $\mathcal{P}_{A_1, l'}$ at time $t'$. 
Both can be expressed in the loop model framework by connecting the ancilla worldline $\gamma_{A,1}$ to bulk worldlines $\gamma_l(t)$ and $\gamma_{l'}(t')$. 
Between times $t,t'$ this is equivalent to drawing a direct loop between $\gamma_l(t)$ and $\gamma_{l'}(t')$, i.e., replacing the ancilla loop by a space-time non-local swap $\calR_{(l,t),(l',t')}$. 
Any two consecutive, single-Majorana measurements $\gamma_{l_n}, \gamma_{l_{n+1}}$ at times $t_n<t_{n+1}$ can thus be replaced by a swap $\calR_{(l_n,t_n), (l_{n+1},t_{n+1})}$. 
By this procedure, the ambiguity in the bulk parity persists only along the loop connecting the first and the final parity breaking measurement via the ancilla. 
After the final parity violating measurement, the ancilla is removed by applying another Hadamard gate and then measuring $Z_A$.
In the bulk, this implements a final swap $\tilde\calR_{(l_N,t_N), (l_{1},t_1)}$, which pushes the parity ambiguity into the qubit represented by the pair $\gamma_{l_1}\gamma_{l_N}$.

When measuring local Pauli operators, e.g., $X$ or $ZX$, the effect of both the ancilla of JW strings need to be combined. 
Each measurement then results in a superposition of all possible loop configurations, including long-range swaps, which are compatible with the state in the circuit, e.g., as in Eq.~\eqref{eq:superposition}. 
In a short loop phase, this again yields only a few possibilities so that the main effects of parity breaking measurements are orientability breaking and long-range swaps. 
By contrast, in a phase with extended loops, parity breaking measurements generate superpositions of many worldline configurations, leading to a rapid growth of entanglement.

\section{Loop Model Phases in Measurement-Only Circuits}\label{sec:measurement_only_ladders}
We now turn our attention toward a family of measurement-only circuits, based on (next-) nearest neighbor qubit measurements. 
We show that such circuits realize phases and phase transitions that are described by Majorana loop models. 
In each case our starting point is a local, Gaussian circuit. 
We distinguish orientable and non-orientable circuits based on bipartiteness of the underlying Majorana graph, which is equivalent to measuring time-reversal symmetric operators.
When orientable, the Gaussian circuits realize loop models corresponding to variants of the $1$-state Potts model, and critical behavior corresponding to the 2D bond percolation universality class. 
In contrast, non-orientable Gaussian circuits are described by the CPLC and feature a transition into the symmetry broken Goldstone phase. 

We will then add non-Gaussian measurements that may or may not preserve the symmetry (i.e. orientability) of the underlying Gaussian circuit. 
In the orientable case, we observe the emergence of a family of loop models, which are connected to the TL or Brauer algebra of (coupled) Potts models. 
Conversely, in the non-orientable case, we find a robust CPLC Goldstone phase: it appears for any type of non-orientable circuit as soon as orientability is broken sufficiently strongly. 

\begin{figure*}[!ht]
    \centering
    \includegraphics[width=\textwidth]{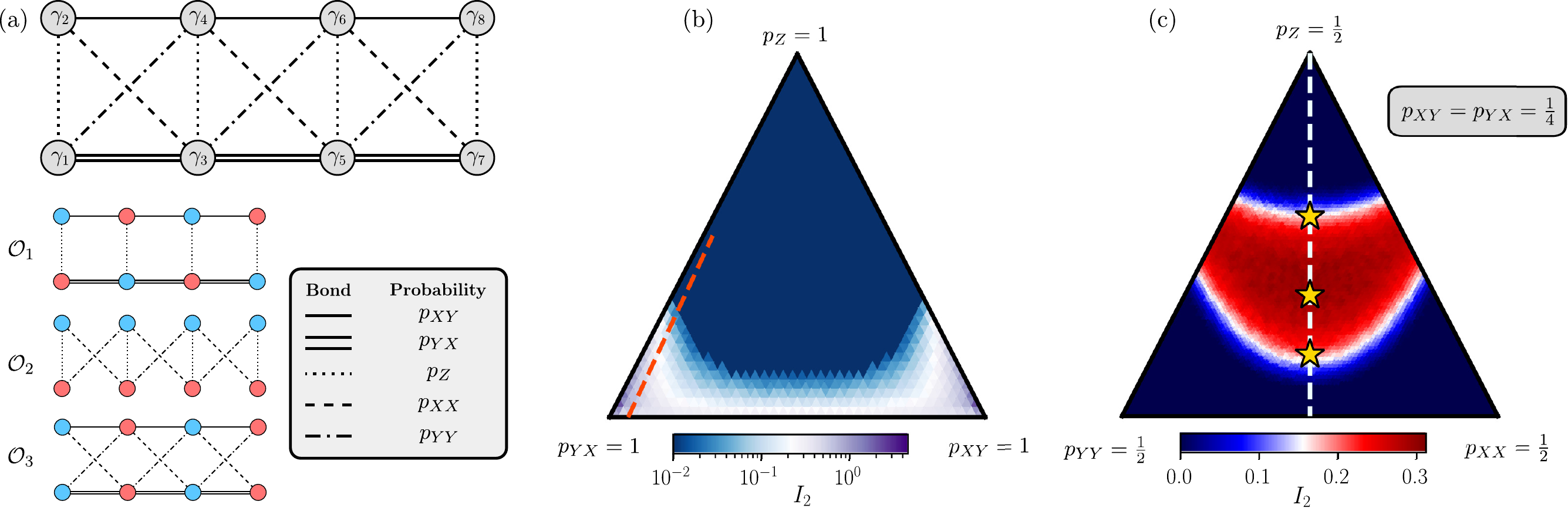}
    \caption{
    \textbf{Loop model for Gaussian Majorana circuits.}
    (a) Each allowed two-fermion parity check is represented by a bond on the lattice. 
    The probability for each bond to be measured and the corresponding Pauli operator are shown in the table.
    Each orientable set $\mathcal{O}_{1,2,3}$ yields a different bipartition of the lattice.
    (b) Phase diagram of the $\mathcal{O}_1$-ladder with $p_Z+p_{XY}+p_{YZ}=1$ for system size $L=1024$.
    For $p_Z>0$ the circuit is in an area law state, which is highlighted by the rapid vanishing of entanglement on a logarithmic scale.
    At large anisotropy ($|p_{XY}-p_{YZ}|\approx1$), rare $Z$-measurements induce a peculiar state with exponentially large correlation length, i.e., a strongly enhanced mutual information, and tunable correlations between the two legs of the ladder.
    The dashed red line along $q=0.05$ corresponds to the parameter interval examined in Fig.~\ref{fig:two_leg_bipartite}.
    (c) Phase diagram of a non-orientable circuit in the $p_{XX} + p_{YY} + p_{Z} = \tfrac12$ plane at fixed $p_{XY} = p_{YX} = \tfrac14$, given by the two-interval mutual information $I_2$ for system size $L=4096$.
    The extended phase with finite $I_2$ is the Goldstone phase.
    The white dashed line gives the $p_{XX} = p_{YY}$ cut used in Fig.~\ref{fig:two_leg_ladder_results}.
    The marked points give exemplary points which are either on the critical line or deep within the Goldstone phase.
    }
    \label{fig:two_leg_ladder}
\end{figure*}

Restricting the measured operators to single or nearest-neighbor qubits and yields exactly five types of Pauli operators that are Majorana bilinears: $\mathcal{O}=\{Z_l, X_lX_{l+1}, Y_lY_{l+1}, X_lY_{l+1}, Y_lX_{l+1}\}$.
It is then convenient to represent the corresponding Majorana circuit on a weighted graph, see Fig.~\ref{fig:two_leg_ladder}a. 
Each vertex $\gamma_l$ of the graph corresponds to a Majorana fermion, and each bond $\hat{O} = i\gamma_l\gamma_m$ specifies a parity-check, measured with probability $p_O$.
For the set of Paulis $\mathcal{O}$ given above, this graph has the structure of a two-leg ladder, shown in Fig.~\ref{fig:two_leg_ladder}a. 
We consider a translationally invariant circuit such that any two identical bonds are measured with the same probability. 
In each temporal ``layer'' of the circuit we employ a \emph{diluted} brickwall structure.
For each brick we either perform a measurement with probability $p = \left(\delta t\right)^{-1}$ or apply the identity with probability $1-p$.
If a measurement is performed, the corresponding Pauli operator $\hat O$ is chosen with probability $p_O$. 
Throughout this paper we take $p = L^{-1}$ such that a unit timestep $\Delta t = 1$ in the circuit corresponds to $L$ randomly selected measurements at randomly selected positions.
When discussing the transfer matrix for a circuit, we implicitly refer to the corresponding brickwall circuit with $p=1$, which should not affect the universal properties of the circuit~\footnote{In the non-orientable circuit, i.e., the CPLC, the identity operation is required in order to realize a robust Goldstone phase. In the orientable case, where we apply transfer matrix methods, no such requirement is present.}.


\begin{figure}[t]
    \centering
    \includegraphics[width=\columnwidth]{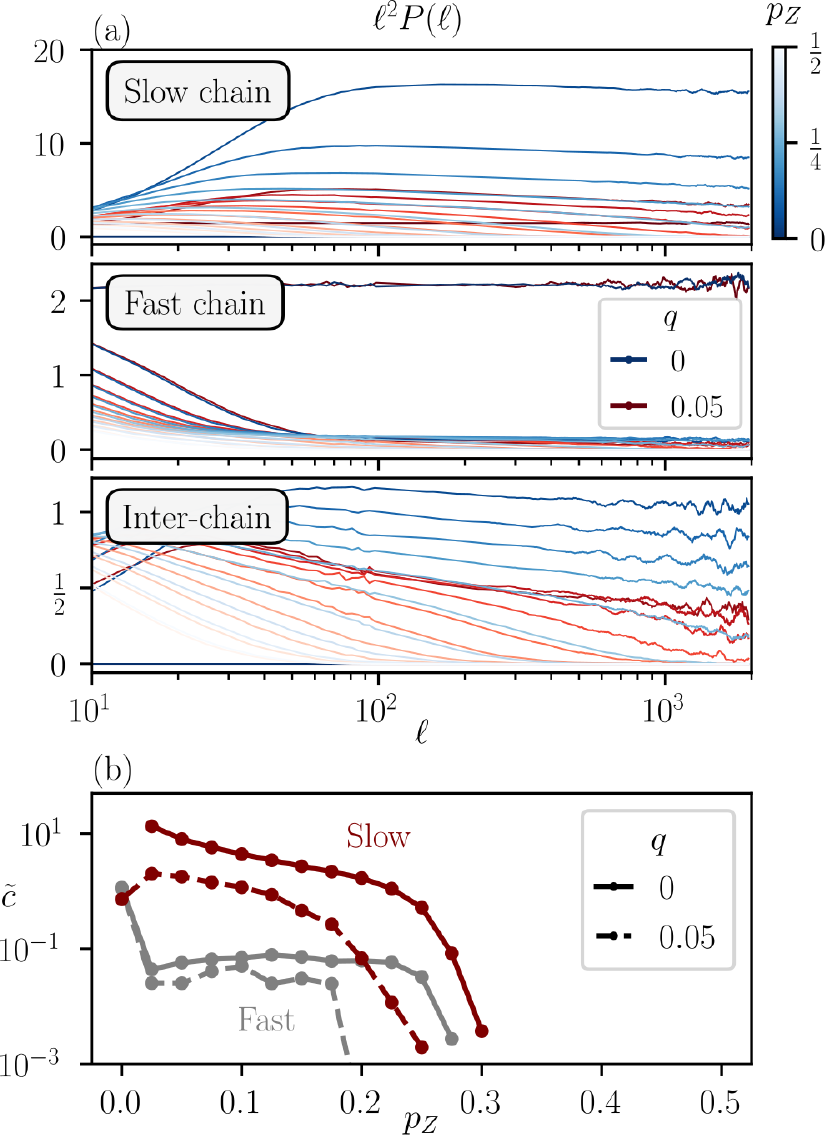}
    \caption{
    \textbf{Measurement-altered criticality for orientable Gaussian circuits: }
    Taking large anisotropy $q = \min\{p_{XY}, p_{YX}\} / (p_{XY} + p_{YX})$, a small but finite inter-chain coupling via $p_Z \neq 0$ alters the criticality of otherwise decoupled chains.
    (a) Length distributions for fermionic loops restricted to either chain or spanning between the two chains.
    The slow and fast chains see a strong enhancement and suppression, respectively, of long loop probability at finite coupling $p_Z$.
    At large $\ell$, $\ell^2 P(\ell)$ approaches a constant when $q=0$ (blue) and decays very slowly at $q=0.05$ (red) in the slow chain, corresponding to approximately critical entanglement induced via the coupling.
    For weak coupling, the fast chain also exhibits almost constant $\ell^2 P(\ell)$, indicating that criticality was modified but not destroyed.
    The inter-chain loop length distribution shows that finite entanglement is shared between the chains at long distances.
    (b) The effective log-law coefficient $\tilde{c}$ in each chain extracted from the asymptotic value of $\ell^2 P(\ell)$ at fixed system size $L$.
    This reflects the enhanced magnitude of $\tilde{c}$ in the slow chain at weak coupling and the reduced but finite criticality of the fast chain.
    }
    \label{fig:two_leg_bipartite}
\end{figure}

\subsection{Orientable Gaussian Circuits}
\label{ss:two_leg_ladder}

From the set of allowed Pauli operators, one can form three translationally invariant and orientable subsets:
\begin{align*}
    \mathcal{O}_1&=\{Z_l, X_lY_{l+1}, Y_lX_{l+1}\},\\
    \mathcal{O}_2&=\{Z_l, X_lX_{l+1}, Y_lY_{l+1}\},\\
    \mathcal{O}_3&=\{X_lY_{l+1}, Y_lX_{l+1}, X_lX_{l+1}, Y_lY_{l+1}\}.
\end{align*}
The corresponding bipartite Majorana graphs are shown in Fig.~\ref{fig:two_leg_ladder}a. 
It is convenient to define two global unitary transformations $U_{Z} \equiv \exp\left(i\tfrac{\pi}{4} \sum_l Z_{l}\right)$, $U_{Z,\text{odd}} \equiv \exp\left(i\tfrac{\pi}{4} \sum_l Z_{2l-1}\right)$, consisting of a $\tfrac{\pi}{2}$-rotation of each or each odd Pauli-$Z$ (i.e. a product of the $\calR_{2l-1, 2l}$ operators).
Application of $U_{Z,\text{odd}}$ maps $\mathcal{O}_1\leftrightarrow \mathcal{O}_2$, making the two circuits identical upon parity sign flips, and it leaves $\mathcal{O}_3$ invariant. 
Application of $U_{Z}$ leaves all three subsets invariant.

\subsubsection{Potts models and percolation}

Each orientable subset $\mathcal{O}_{1,2,3}$ contains at least one further subset of operators whose projectors form a TL algebra. 
For instance, the set of measurements $\{Z_l, X_lX_{l+1}\}=\{i\gamma_{2l-1}\gamma_{2l}, i\gamma_{2l}\gamma_{2l+1}\}$ forms the TL algebra generated by $\{b_l = \calP_{l,l+1}\}$. 
The evolution of the circuit is then described by the transfer matrix
\be
    \begin{aligned}
        T &= T_1 T_3 \dots T_{L-1} T_2 T_4 \dots T_L, \\ 
        T_{2l-1} &= 1 + p_Z b_{2l-1}, \\
        T_{2l} &= 1  + p_{XX} b_{2l}
    \end{aligned}
\ee
which is identical to the transfer matrix in the 2D 1-state Potts model. 
Depending on the probabilities $p_Z, p_{XX}$, the odd ($b_{2l-1}$) or even $(b_{2l}$) parity measurements dominate. 
This yields a dimerized ground state in the Potts model given by a configuration of short loops over even or odd bonds. 
The latter is equivalent to a stationary state in the circuit with dimerized Majorana parities that features area law entanglement and zero (one) pairs of Majorana edge states for $p_Z>p_{XX}$ ($p_Z<p_{XX}$). 

The transition between the two topologically inequivalent area law states at $p_Z=p_{XX}$ is described by the critical 1-state Potts model, corresponding to the universality class for 2D bond percolation.
In the loop framework, this yields an algebraic distribution of loop lengths $\ell$, $P(\ell)=3\log(2)\tilde{c} \ell^{-2}$ and logarithmically growing entanglement entropy $S_L \sim \tfrac{\tilde c}{3}\log(L)$ with $\tilde{c} = \tfrac{3\sqrt{3}\log(2)}{2\pi}$~\cite{Cardy_2000, Jacobsen_2008} (see Table~\ref{tab:ref_table}).
The universality of this transition has been confirmed in previous works, where measurements of the set $\{Z_l, X_{l}X_{l+1}\}$ were motivated as a \emph{projective Ising model}~\cite{Roser_2023, Lang_2020} or a measurement-only version of the \emph{repetition code}~\cite{Li_2021_Z2, Nahum_2020}.
Applying $U_{Z,\text{odd}}$, $U_{Z}$ or their product to $\{Z_l, X_lX_{l+1}\}$ yields three further sets of measurements that realize the same TL algebra and thus the same universal behavior. 
Since both $U_{Z}$ and $U_{Z,\textrm{odd}}$ correspond to constant-depth local unitary circuits, the topology of the two area-law states is preserved --- local swaps transform the dimerization pattern but do not add or remove edge modes.

Two further \emph{independent} TL algebras $\{d_l = \calP_{2l-1, 2l+1}\}, \{f_l = \calP_{2l, 2l+2}\}$ are realized by the sets $\{Y_lX_{l+1}\}=\{i\gamma_{2l-1}\gamma_{2l+1}\}$ and $\{X_lY_{l+1}\}=\{i\gamma_{2l}\gamma_{2l+2}\}$. 
Application of $U_{Z}$ exchanges $\{d_l\}\leftrightarrow\{f_l\}$, while under $U_{Z,\text{odd}}$, $\{Y_lX_{l+1}\}\leftrightarrow\{X_{2l-1}X_{2l}, Y_{2l}Y_{2l+1}\}$ and $\{X_lY_{l+1}\}\leftrightarrow\{Y_{2l-1}Y_{2l}, X_{2l}X_{2l+1}\}$. 
Together, the two sets realize two \emph{independent} 1-state Potts models. 
While in principle the parameters for these Potts models can be tuned independently, translational invariance of the measurement probabilities yields an additional constraint. 
For the set $\{Y_lX_{l+1}, X_{l}Y_{l+1}\}$, translational invariance places both Potts models exactly at the critical point~\footnote{This can be made more apparent by a unitary transformation $\prod_l e^{-i\pi X_{2l-1} X_{2l} / 4}$: it maps the circuit to a variant of the cluster-model~\cite{Lavasani_2021}, whereas here $XZX$ and $Z$ are measured with equal probabilities.} but with distinct timescales fixed by $p_{XY}$ and $p_{YX}$.
In contrast, for the set $\{X_lX_{l+1}, Y_{l}Y_{l+1}\}$ both Potts models are identical but can be tuned through different phases. 

Once we extend the measurements to include any full subset $\mathcal{O}_{1,2,3}$, the circuit remains orientable but is no longer described by uncoupled TL algebras. 
Instead, the Potts models are coupled by a relevant, Gaussian operator. Below, we consider one particular case of such coupled Potts models in the Gaussian circuit.

\subsubsection{Apparent criticality altered by measurement}

Consider the orientable set $\mathcal{O}_1$. 
It realizes two critical 1-state Potts models, represented by the chain of either the even $\{X_lY_{l+1}\}$ or the odd $\{Y_{l}X_{l+1}\}$ Majorana modes. 
The two are coupled by measurements of $\{Z_l\}$.
This Gaussian term is relevant and introduces a time- and length scale, pushing the circuit into an area law phase.
However, when both critical Potts models have significantly different characteristic timescales, i.e., for strong anisotropy between the measurement probabilities, the correlation length remains exponentially large, c.f., Fig.~\ref{fig:two_leg_ladder}b. 
This yields an intriguing dynamical regime:
A circuit of finite size or depth displays long-range correlations and apparent critical behavior, up to exponentially large distances, which is tunable by $\{Z_l\}$-measurements -- a general feature when Potts models are weakly coupled by a relevant operator.

Let us define $q = \min\{p_{YX},p_{YX}\} / (p_{XY} + p_{YX})$, which measures the anisotropy between the two Potts models. 
At strong anisotropy and weak coupling, i.e., $q, p_Z \ll 1$, the typical timescale on which each of the two Potts models evolves is $\approx q^{-1}, (1-q)^{-1}$. 
Weak coupling between the two models then provides an additional channel for the \emph{slow} Potts model to spread correlations via the \emph{fast} one. 
Consider for instance the case $p_{YX} \ll p_{XY}$ such that odd Majoranas $\{\gamma_{2l-1}\}$ evolve much faster than even ones $\{\gamma_{2l}\}$. 
In order for a pair $\gamma_{2l}\gamma_{2l+2m}$ on the slow chain to become entangled, a loop endpoint must travel $m$ sites.
This can be accomplished either by waiting $\ge q^{-m}$ layers for rare, even-mode measurements or by taking the fast-lane, harnessing frequent odd-mode measurements to travel the same distance in $\ge (1-q))^{-m} p_Z^{-2}$ layers.
At long distances $m \gtrsim 2\log(p_Z) \log(q / (1-q))$, the latter process dominates. 
Correlations in the slow chain are then significantly enhanced by the weak coupling. 
Hence, the fast (slow) Potts model acts as bath for the slow (fast) one, enhancing (reducing) its tendency to form long loops $\ell \gg 1$.
A similar scenario has been observed in other monitored quantum systems, in particular when coupling two critical Ising chains by measurements~\cite{Murciano_2023}. 

The modification of the length distribution $P(\ell)$ for loops within and between the two Majorana chains is seen in Fig.~\ref{fig:two_leg_bipartite}. 
Both intra-chain distributions converge to an approximate power-law $\ell^2 P(\ell) \to \alpha_\ell$, whereas $\alpha_\ell$ decays very slowly with $\ell$ on a logarithmic scale. 
This reflects the exponentially large correlation length $\xi$, such that for loop lengths $\xi>\ell\gg1$ a power-law behavior, reminiscent of a critical state, is observed. 
As a consequence, each individual chain displays a logarithmic growth of the entanglement entropy at finite system size. 
For the slow (fast) chain, the magnitude of the probability density is enhanced (reduced) by roughly one order of magnitude. 
Long loops are almost exclusively accumulated in the slow chain and, to a lesser  extent, between the chains.
From the asymptotic value $\lim_{\xi>\ell\gg1}\ell^2 P(\ell)=\tfrac{\tilde c}{3}$, we can extract the coefficient $\tilde{c}$ for the logarithmic growth of the intra-chain entanglement entropy $S_{L/2}=\tfrac{\tilde c}{3}\log(L)$, shown in Fig.~\ref{fig:two_leg_bipartite}b. 
This value is no longer universal.
Instead it can be tuned to arbitrary values by adjusting the measurement probabilities $p_Z,q$ and thereby the anisotropy between the Potts models and their coupling strength. 
Since the correlation length $\xi$ diverges for $p_Z \to 0$, this provides a knob to control both the amount of long-range correlated loops in each individual chain as well as the length scale up to which the apparent critical behavior can be observed. 

We note that similar scenarios have emerged in related, orientable loop models; earlier works on coupled network models for the spin quantum Hall effect have shown that coupling two critical 1-state Potts models via a Gaussian operator immediately opens a gap, leading to a finite but large correlation length~\cite{Beamond_2003, Chalker_2011}.
Furthermore, the orientable boundary of the CPLC phase diagram studied in Ref.~\cite{Nahum_2013} gives rise to a coupled Potts model interpretation, for which the correlation length grows exponentially large with the tuning parameter.

\subsection{Orientable non-Gaussian Circuits}\label{sec:orient_non_gauss}

Let us now consider measuring the orientable subset $\mathcal{O}_3=\{X_lX_{l+1}, Y_lY_{l+1}, Y_lX_{l+1}, X_lY_{l+1}\}$ and the non-Gaussian operators $\{Z_lZ_{l+1}, X_lX_{l+2}\}$.
The latter are compatible with orientability but introduce 8-leg vertices into the loop diagram. 
This prohibits a precise microscopic mapping to a TL algebra for the circuit.
Nonetheless, we can construct an effective transfer matrix based on the TL algebra of the Gaussian operators, and we show that it accurately describes the dynamics in the circuit.
The structure of the transfer matrix is again familiar from the statistical mechanics of coupled $Q$-state Potts models~\cite{Fendley_2008}.
We argue that this approach is general for both Gaussian and non-Gaussian orientable circuits and thus establishes close contact between one-dimensional measurement-only circuits and 2D Potts models.  

Adding symmetry-preserving, non-Gaussian measurements is an irrelevant perturbation to an extended phase of Gaussian measurements, though it may become relevant in the vicinity of a Gaussian critical point. 
Furthermore, when non-Gaussian measurements become the dominant generator of evolution, they may drive the circuit towards a strong coupling fixed point.
In the following, we discuss three general examples for this scenario: 
(i) by adding non-Gaussian measurements in the vicinity of the percolation critical point of the $1$-state Potts model, we split the critical point into two BKT transitions, which enclose an emergent critical phase, 
(ii) when inducing additional frustration by measuring incommensurate non-Gaussian operators, the critical phase transforms into a volume law entangled phase and
(iii) when non-Gaussian measurements dominate, they give rise to a strong coupling tricritical point with an emergent Ising universality. 
We show how (i)-(iii) can be understood from an elementary version of the problem admitting an exact algebraic description.

\begin{figure*}[t]
    \centering
    \includegraphics[width=\textwidth]{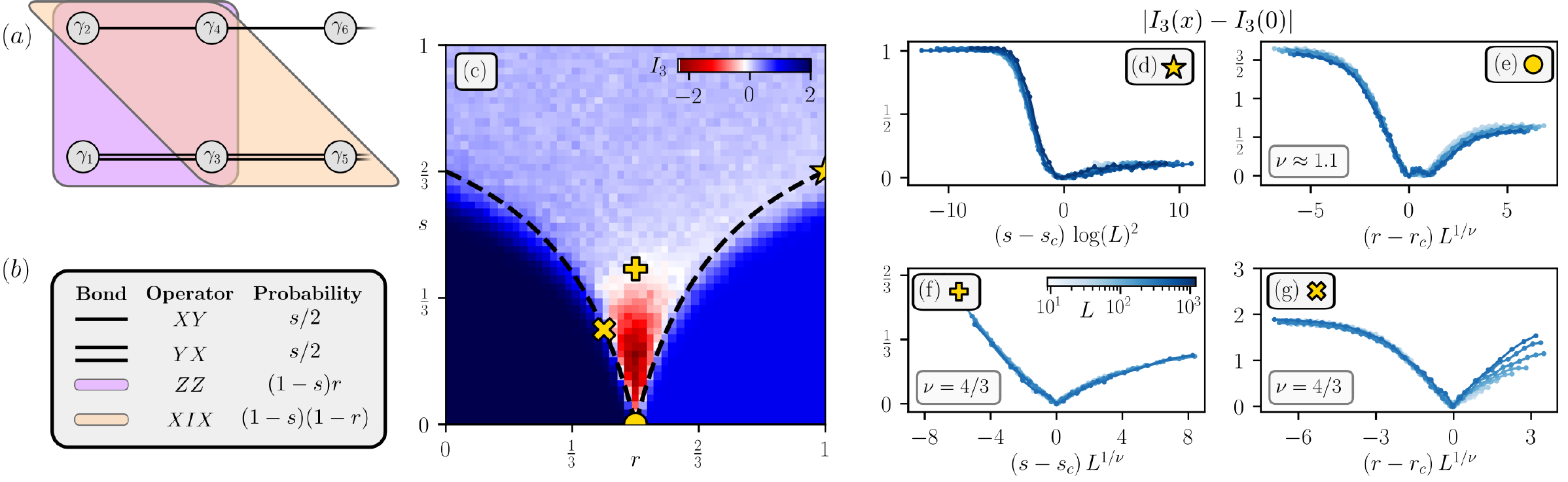}
    \caption{
    \textbf{Orientable non-Gaussian circuit:}
    (a) Majorana lattice for the two-leg ladder with four-fermion measurements.
    Shaded plaquettes denote $Z_lZ_{l+1}$ (purple) and $X_lX_{l+2}$ (orange) measurements.
    (b) Probabilities for the allowed two and four-fermion measurements.
    Quantity $s \in [0,1]$ sets the relative probability for two-fermion versus four-fermion measurements, while $r$ determines which type of four-fermion terms are measured.
    (c) Phase diagram in the $r$-$s$ plane showing three distinct phases: an area-law phase at strong 4-fermion measurement rate, a critical phase extending from the non-interacting limit, and a volume-law phase arising from frustration between 4-fermion terms.
    The dashed black line marks the analytic expression for the critical point $s_c = 2\abs{2r-1}/(1 + 2\abs{2r-1})$.
    The apparent deviation of the data from expected critical line results from finite size effects which are particularly strong for larger $\abs{r - \tfrac12}$, where the BKT transition gives logarithmically slow convergence to the thermodynamic limit.
    (d-g) Critical rescaling of the tripartite mutual information $I_3$ along various cuts in the $r$-$s$ plane.
    In all plots, the critical value $I_3(x=0)$ is subtracted off and data are shown for system sizes up to at least $L=360$.
    (d) When $r = 1$, the critical and area-law phases are separated by a BKT transition at $s_c = \tfrac23$.
    (e) With only interactions ($s=0$), the two area-law phases are separated by a conventional second-order transition at $r_c = \tfrac12$ with critical exponent $\nu\approx1.1$, close to the expected $\nu=1$ for Ising criticality.
    (f) Fixing $r = \tfrac12$, we observe a percolation transition between a critical phase and a volume-law phase with logarithmic corrections.
    (g) A cut at $s=\tfrac14$ reveals a percolation transition with $\nu=\tfrac43$ separating the area- and volume-law phases.
    }
    \label{fig:interacting_twoLeg_decoupled}
\end{figure*}

We consider a simplified version of the circuit, based on measuring the Gaussian operators $\{XY, YX\}$ with probability $\tfrac{s}{2}\equiv p_{XY}=p_{YX}$ and the non-Gaussian operators $Z_lZ_{l+1}$ and $X_lX_{l+2}$ with probabilities $p_{ZZ}=(1-s)r$ and $p_{XIX}=(1-s)(1-r)$, respectively.
The phase diagram in Fig.~\ref{fig:interacting_twoLeg_decoupled}c shows the different transitions realizing scenarios (i)-(iii). 
Here the choice of Gaussian operators admits a pair of TL algebras with generators $e_l \equiv \calP_{2l-1,2l+1} = \tfrac12 (\mathds{1} + X_lY_{l+1})$ and $f_l \equiv \calP_{2l,2l+2} = \tfrac12(\mathds{1} + Y_lX_{l+1})$, respectively.
The transfer matrix for the Gaussian circuit then is equivalent to that of two uncoupled, critical 1-state Potts models.
Treating the non-Gaussian operators as a perturbation, the transfer matrix $T$ for the interacting circuit can be expressed in terms of the original TL generators (see App.~\ref{app:effective_transfer_matrix})
\begin{align}
    T&= \sum_i \left[\lambda (e_i + f_i) + e_i\left((1+\delta r)f_i + (1-\delta r)f_{i+1} \right) \right],\label{eq:effective_T}\\
    \lambda &\equiv \tfrac{s}{2(1-s)}-|\delta r|, \ \delta r \equiv 2r-1.\nonumber
\end{align}
We will now go through the scenarios (i)-(iii) in the phase diagram with the help of the transfer matrix $T$.

\subsubsection{Emergent critical phase and splitting of the percolation critical point into two BKT transitions} 

Consider the limit $s\to 1$ and $\delta r=\pm 1$, which corresponds to two $1$-state Potts models, which are tuned towards their critical point. 
Each is described by a TL algebra $\{e_i\}, \{f_i\}$. 
Then adding measurements of either $\{Z_lZ_{l+1}\}$ or $\{X_lX_{l+2}\}$ introduces the couplings $e_i f_i$ or $e_i f_{i+1}$, which both generate yet another TL algebra for a single, critical $1$-state Potts model. 
This coupling is known to be an irrelevant perturbation~\cite{Fendley_2008} and thus for $\lambda>0$ the critical Gaussian state remains robust. 
Increasing the rate of non-Gaussian measurements, the circuit reaches a strong-coupling fixed point at $\lambda=0$, i.e., at a critical $s = s_c = \tfrac23$.
Here, the two critical $1$-state Potts models combine into a single emergent one. 
For $\lambda<0$, i.e., $s<s_c$, non-Gaussian measurements dominate and induce an area law phase, whose topological properties depends on the sign of $\delta r$.
With $\delta r=1$ the area law phase is trivial, whereas for $\delta r=-1$ one pair of edge states $\sim i\gamma_1\gamma_{2L}$ is stabilized, see Fig.~\ref{fig:interacting_twoLeg_decoupled}a. 
In both cases, the phase transition in the coupled Potts models follows a BKT scenario, which is visualized by the finite size scaling collapse of the mutual information in Fig.~\ref{fig:interacting_twoLeg_decoupled}d.

When moving away from the limit $\delta r=\pm1$, both $e_if_i$ and $e_if_{i+1}$ appear simultaneously, reflecting the competition between the two types of non-Gaussian measurements. 
For any $\delta r$, the dominant non-Gaussian term induces a robust area law phase for $\lambda<0$, which becomes unstable for $\lambda\ge0$. 
This yields a precise analytical estimate for the phase boundary at $s=s_c(\delta r)=\tfrac{2|\delta r|}{1+2|\delta r|}$, which is confirmed in Fig.~\ref{fig:interacting_twoLeg_decoupled}c. 
For large anisotropy $\delta r$ between the two couplings $e_if_i, e_if_{i+1}$ or for large $\lambda>0$ the Gaussian critical phase remains robust. 
This is accompanied by a BKT transition into the area law phase at critical $s_c$. 
Numerical simulations of the circuit witness an extended critical phase for $\tfrac{1}{3} \lesssim s$ and correspondingly a BKT transition for $\tfrac{1}{4}\lesssim|\delta r|$.

The emergence of a robust critical phase, separated by a BKT critical line from an area law regime, appears to be generic when symmetry-preserving non-Gaussian measurements are added to the sets $\mathcal{O}_{1,2,3}$. 
The prerequisite is that the unperturbed circuit is in the vicinity of the $1$-state Potts model critical point. 
We show in App.~\ref{sec:Broadening} that non-Gaussian measurements cause the single $1$-state Potts model critical point to branch out into two separate BKT transitions, which enclose an extended critical phase. 
This might be understood as follows. 
When an orientable set $\mathcal{O}_{1,2,3}$ is pushed to a percolation critical point, it can be decomposed into two $1$-state Potts models, one of which is critical. 
Non-Gaussian measurements then couple both Potts models into a robust critical phase. The scenario in Fig.~\ref{fig:interacting_twoLeg_decoupled} represents one side of the broadened transition between two area law phases.
A similar picture has been observed for orientable Gaussian measurements, which at the critical point are perturbed by unitary gates~\cite{bao2021symmetry}.

\subsubsection{Tricritical strong coupling fixed point: Emergent Ising universality}\label{ss:ising}

The two critical lines with $\lambda=0$ meet at the critical point $(s,\delta r)=(0,0)$, see Fig.~\ref{fig:interacting_twoLeg_decoupled}c. 
The merging of two critical lines is typical for a fine-tuned tricritical point, which then realizes a different universality class. 
Indeed, it has been shown that the tricritical point of the ferromagnetic $Q$-state Potts model with $\sqrt{Q} = 2\cos(\pi / k)$ is related to the critical point of the ferromagnetic $Q'$-state Potts model, with $\sqrt{Q'} = 2\cos(\pi/(k+1))$, via the so-called $\epsilon$-$\eta$ duality~\cite{Delfino_2017, Duplantier_1989, Duplantier_1987, Janke_2004, Nienhuis_1979_RG}.
We argued above that each critical line at $\lambda=0, \delta r\neq 0$ corresponds to a single, critical $1$-state Potts model, i.e., $k=3$. 
The duality then predicts the tricritical point at $(s,\delta r) = (0,0)$ to be described by a critical $2$-state Potts model ($k=4$), i.e., the Ising universality class.
This can be made rigorous by recalling that in the Potts model, four-spin interactions (e.g. the $e_if_i$ coupling terms) are mapped to vacancies under a duality transformation~\cite{Wu_1981, Nienhuis_1982, Coniglio_1982, Knops_1993}. 
This maps the two coupled, dense Potts models to a single, dilute $1$-state Potts model in which $s$ tunes the temperature and $\delta r$ tunes the vacancy density away from the critical point. 
Tuning $(s,\delta r) \rightarrow (0,0)$ takes the model to its tricritical point and the $\epsilon-\eta$ duality yields the critical Ising model.

Numerical simulations confirm the correlation length critical exponent $\nu\approx 1$ of the Ising universality class (see Fig.~\ref{fig:interacting_twoLeg_decoupled}c).
In order to further support the emergent Ising universality, we extract the bulk and surface critical behavior through coupling the circuit to ancilla (see Appendix~\ref{app:ising_transition}).
This provides information on out-of-time ordered correlation functions and yields the critical exponents $\eta\approx 0.24$ in the bulk and $\eta_\parallel\approx 1.1$ at the surface.
Both are compatible to the Ising values $\eta = 1/4$ and $\eta_\parallel = 1$~\cite{Cardy_1984}. 
Furthermore, $\eta_\parallel = 1$ is consistent with the scaling of the mutual information, which was obtained in Ref.~\cite{Lin_2023} for $ZIZ$ and $XX$ measurements.

We note three peculiar points: 
(i) Replacing the set $\{X_lX_{l+2}, Z_lZ_{l+1}\}$ by $\{X_lZ_{l+1}X_{l+2}, Z_lZ_{l+1}\}$ yields an identical measurement-frustration graph yet provides a more intuitive picture of the tricritical $1$-state Potts model. 
The Gaussian set $\{X_lZ_{l+1}X_{l+2}\}$ realizes two identical Majorana chains, which are both coupled by $\{Z_lZ_{l+1}\}$ measurements and simultaneously pushed to the critical point. 
(ii) The same frustration graph is provided by the set $\{X_lX_{l+2}X_{l+3}X_{l+4}, Z_l\}$, which is the measurement-only version of an Ising chain with multi-spin interactions.
The corresponding Hamiltonian has an 8-fold degenerate ground state and is related to the 8-state Potts model~\cite{Alcaraz_1986, Penson_1982, Milsted_2015, Rahmani_2015a, Rahmani_2015b, Blote_1986, Selke_1988, Turban_2016, OBrien_2018}, which undergoes a first-order phase transition. 
This highlights the observed trend: orientable measurement-only circuits realize $Q$-state Potts models, for which $Q$ is reduced compared to their Hamiltonian counterparts. 
(iii) Despite the critical exponents determining the Ising universality class, we observe an intriguing behavior of the entanglement entropy at the critical point. 
It grows logarithmically $S_{L/2}=\tfrac{\tilde c}{3}\log(L)$ with prefactor $\tilde{c} = 2$. 
This is significantly larger than the expected value $\tilde{c}_{Q=2} = (1 + \sqrt{2})\log(2)/\pi \approx 0.53$ from the loop distribution of the $2$-state Potts model~\cite{Jacobsen_2008} (see Table~\ref{tab:ref_table} and Appendix~\ref{app:ising_transition}).
This discrepancy may result from a difference in boundary CFTs between the critical Ising model and the tricritical 1-state Potts model.

\subsubsection{Emerging volume law from competing non-Gaussian measurements} 

For small anisotropy $\abs{\delta r} \ll 1$ and positive $\lambda>0$, all the terms in the transfer matrix in Eq.~\eqref{eq:effective_T} become relevant and none can be treated as a small perturbation. 
In this regime, the critical phase of the two $1$-state Potts models for $\lambda>0$ becomes unstable and the circuit enters a volume law entangled state. 
The volume law is signalled by a negative value of the tripartite information $I_3$ which scales with system size~\cite{Gullans_2020_probe, Zabalo_2020}, shown in Fig.~\ref{fig:interacting_twoLeg_decoupled}c). 
The volume law phase is symmetric around $\delta r=0$.
It is separated from the critical phase of the Gaussian circuit by a critical line at $\tilde s_{c}(r)\approx\tfrac13$ and from the area-law phases by the already established critical line $\lambda=0$.
At $(s,\delta r)=(0,0)$ it terminates at the tricritical point.
As a consequence, two additional tricritical points emerge at $(s,\delta r) \approx (\tfrac13, \pm \tfrac14)$ where all three types of entanglement phases (area law, volume law, critical) meet.
These latter tricritical points are, however, difficult to locate precisely.
Along each boundary of the volume law phase, except for the tricritical points, we find the critical scaling behavior of the $1$-state Potts model, i.e., percolation, see Fig.~\ref{fig:interacting_twoLeg_decoupled}f,g. 

While a percolation transition separating a volume-law phase from an area-law phase has been observed previously in measurement-only~\cite{Ippoliti_2021} and monitored unitary circuits~\cite{Li_2019,Nahum_2017} a percolation transition separating the volume-law and critical phases is unconventional. 
We observe that the logarithmic growth of the entanglement entropy at this transition remains largely unaltered, while a volume-law contribution emerges on top. 
A possible explanation of this scenario and of the scaling behavior at the transition might be that the critical sector is decoupled from an otherwise area-law to volume-law transition.

The emergence of the volume law entangled phase can be explained from two complementary views in the Potts model framework. 
When starting from the BMW algebra for strongly coupled Potts models, the operator $e_i + f_i$ acts as a perturbation that generates line-crossings~\cite{Fendley_2008}, i.e., emergent swap operations $\calR_{\text{eff}}$.
For $\abs{\lambda / \delta r} \sim 1$ and $\lambda>0$, this yields an effective circuit implementing frequent swap operations, which rapidly generate long, orientable loops and thus extensive entanglement. 
For $\lambda < 0$ the same types of swaps arise but the negative sign results in the cancellation of many trajectories in the partition function, suppressing the mobility of the worldlines.

In turn, when starting from the tricritical point at $(s,\delta r)=(0,0)$, both $\delta r \neq 0$ and $s > 0$ correspond to relevant perturbations, as explained above~\cite{Nienhuis_1982}.
The imbalance parameter $\delta r \neq 0$ tunes the chemical potential for vacancies in the dual dilute model and drives the system toward an ordered area-law fixed point. 
Tuning $s>0$ reintroduces the generators $e_i, f_i$ of the original $1$-state Potts models. 
In the dual framework, taking finite $s$ corresponds to tuning the temperature and pushing the $2$-state Potts model into its high temperature phase.
This removes the energetic penalty for forming long loops in the bulk and yields a disordered, volume law entangled state at the temporal boundary of the circuit.

\subsection{Non-Orientable Gaussian Circuits}\label{sec:non_orient_Gauss}

Once the set of measured Gaussian operators no longer admits a bipartition of the Majorana graph, e.g., the union of $\mathcal{O}_{1,2,3}$, the circuit is no longer orientable. 
The evolution then maps to a loop model which is described by the Brauer algebra of CPLC instead of a TL or walled Brauer algebra. 
For sufficiently strong competition between incompatible measurements, this non-orientable loop model is known to enter an extended long-loop phase, the \emph{Goldstone phase}. 
Both the critical point at the boundary of the Goldstone phase as well as the phase itself display universal scaling behavior, which is described by an $\mathbb{RP}^{n-1}$ NL$\sigma$M~\cite{Nahum_2013}. 
Circuits realizing this Goldstone phase display a peculiar and universal logarithmic correction to entanglement, purification and bulk correlation functions~\cite{Nahum_2013, Loio_2023, Fava_2023}. 

We confirm that non-orientable Gaussian circuits indeed display the universal behavior of CPLC by examining bulk and boundary correlation functions for the following setup. 
We measure the operators $\{XY, YX\}$ with equal probability $p_{XY}=p_{YX}=\tfrac14$ and add measurements of $\{Z, XX, YY\}$ such that $p_Z + p_{XX} + p_{YY} = \tfrac12$. 
The circuit is symmetric for $p_{XX}\leftrightarrow p_{YY}$, i.e., under the global unitary $U_{Z}$. 
The two limits $p_Z=0,\tfrac12$ are both orientable and correspond to two topologically distinct area law phases. 
As shown in Fig.~\ref{fig:two_leg_ladder}c, they are separated by an extended phase with an increased mutual information and entanglement.
By examining a cut along the symmetric point $p_{XX} = p_{YY} = \tfrac14(1-2p_Z) \equiv p$ (white dashed line in Fig.~\ref{fig:two_leg_ladder}c), we show that the entangled phase exhibits all key characteristics of the CPLC Goldstone phase.

\begin{figure}[t!]
    \centering
    \includegraphics[width=0.8\columnwidth]{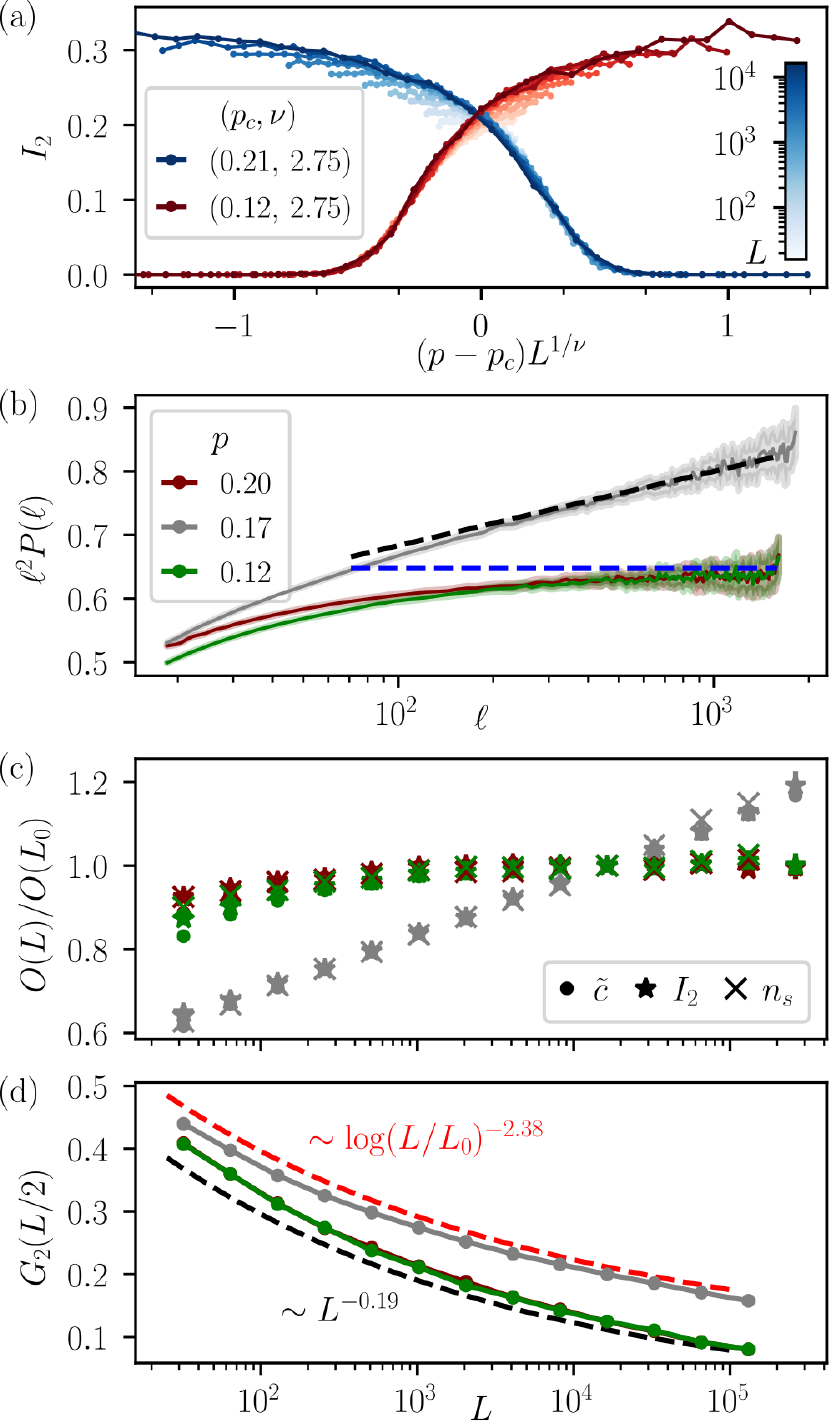}
    \caption{
    \textbf{Non-orientable Gaussian circuit: }
    Data are taken along the line-cut in Fig.~\ref{fig:two_leg_ladder}c with $p_{XY}=p_{YX}=\tfrac14$ and $p\equiv p_{XX} = p_{YY} = \tfrac14(1-2p_Z)$ in the steady state $T=8L$ for system sizes up to $L=2^{17}$.
    (a) Scaling collapse of the mutual information $I_2$ identifies the critical points $p_{c,1} = 0.12$, $p_{c,2} = 0.21$ and a joint critical exponent $\nu\approx 2.75$.
    (b) Loop length distribution $P(\ell)$ normalized by $\ell^2$ to highlight the asymptotic behavior.
    In the Goldstone phase (grey), CPLC predicts a logarithmic correction $\ell^{2}P(\ell) = \tfrac{1}{2\pi^2}\log(\ell)$ (dashed black line).
    By contrast, at the critical points $p = p_{c, 1/2}$, (maroon and green), CPLC predicts a conventional critical scaling $\ell^2P(\ell) \approx 2.035 / \pi$ ~\cite{Nahum_2013} (dashed blue line).
    (c) System-size dependence of the log-law coefficient $\tilde{c}$ (circles), mutual information $I_2$ (stars), and spanning number $n_s$ at circuit depth $T=L$ (crosses), normalized against the value at $L_0 = 2^{14}$.
    In the Goldstone phase (gray) we observe logarithmic corrections, $O(L) \propto \log(L)$, whereas at the transition $p=p_{c,1/2}$ all three quantities approach a system-size independent value.
    (d) The scaling of the two-leg watermelon correlator $G_2(L/2)$ is consistent with CPLC \cite{Nahum_2013}.
    At the transition, there is power-law decay $G_2(L/2) \sim L^{-2x_2}$ with $x_2 \approx 0.096$ (dashed black line), whereas in the Goldstone phase correlations decay slower than algebraic, $G_2(L/2) \sim \log(L/L_0)^{-\alpha_2}$ with $\alpha_2 \approx 2.38 \pm 0.2$.
    }
    \label{fig:two_leg_ladder_results}
\end{figure}

\emph{Correlation length critical exponent:} Two critical points, $p_{c,1} = 0.12$ and $p_{c,2} = 0.21$, separate the Goldstone phase from the area law phases.
At both points, a finite-size scaling collapse of the mutual information $I_2$ (Fig.~\ref{fig:two_leg_ladder_results}a) confirms the CPLC exponent $\nu= 2.75$~\cite{Nahum_2013}.

\emph{Loop length distribution:} Both critical points feature a power law scaling of the loop length distribution $P(\ell)$. 
It approaches a constant $\ell^2 P(\ell)\to\alpha$ with $\alpha\approx 0.64$ in Fig.~\ref{fig:two_leg_ladder_results}b, which is consistent with the value from CPLC: $\alpha= \frac{2.035}{\pi}\approx0.648$ arising from the renormalized spin stiffness \cite{Nahum_2013, Sang_2021}.
In the Goldstone phase, the distribution acquires the characteristic logarithmic correction $\ell^2 P(\ell) \sim \beta \log(\ell)$, with $\beta \approx 0.0565$ being consistent with the CPLC value $\tfrac{1}{2\pi^2} \approx 0.0507$~\cite{Nahum_2013, Sang_2021}, see Table~\ref{tab:ref_table}.

\emph{Entanglement entropy: } We compute the steady-state entanglement entropy $S_A(L)$ of a contiguous subsystem of $|A|$ qubits in a system of size $L$. 
For both the critical points and the entangled phase, we make the ansatz $S_A(L) = \frac{\tilde{c}(L)}{3} \log_2\left[\frac{L}{\pi}\sin\left(\frac{\pi |A|}{L}\right)\right] + \beta$.
The critical points $p_{c,1/2}$ display an expected logarithmic scaling, $\tilde{c}(L), I_2(L) \sim \textrm{const}$, shown in Fig.~\ref{fig:two_leg_ladder_results}c. 
In the Goldstone phase, the logarithmic correction to the loop distribution leads to an enhanced entanglement with $\tilde{c}, I_2 \propto \log(L)$.

The precise form of the entanglement and mutual information follow from the loop length distribution $P(\ell)$. 
Both can be computed through an appropriate integral over $P(\ell)$ in the continuum limit of the lattice (see App.~\ref{app:computation_details} for details).
At the critical points, $P(\ell) = \alpha \ell^{-2}$ yields $I_2 = \alpha \log(4/3)$ and $S_A = \alpha \log(\abs{A}) + \textrm{const}$.
In contrast, the distribution $P(\ell) = \beta\log(\ell / \ell_0) \ell^{-2}$ for cut-off length-scale $\ell_0$ yields $I_2 = \beta\log(4/3) \log(\abs{A}/\ell_0) + \textrm{const}$ (we took $\abs{A} = L/8$) and $S_A = \tfrac12 \beta \left[\log(\abs{A}\right]^2 + \beta \log(\abs{A})(1 - \log(\ell_0)) + \textrm{const}$.
For universal prefactor $\beta = \tfrac{1}{2\pi^2}$ from CPLC, we expect $I_2 \sim \tfrac{\log(4/3)}{2\pi^2}\log(L) \approx 0.0146\log(L)$ and $\tilde{c} \sim \frac{3\log(2)}{2\pi^2} \log(L) \approx 0.105\log(L)$. 
Indeed the data at $p = \tfrac16$ exhibits a logarithmic growth of $I_2$ ($\tilde c$) with prefactor $0.0194$ ($0.0974$), both comparable to the universal CPLC values.

\begin{figure*}[ht]
    \centering
    \includegraphics[width=\textwidth]{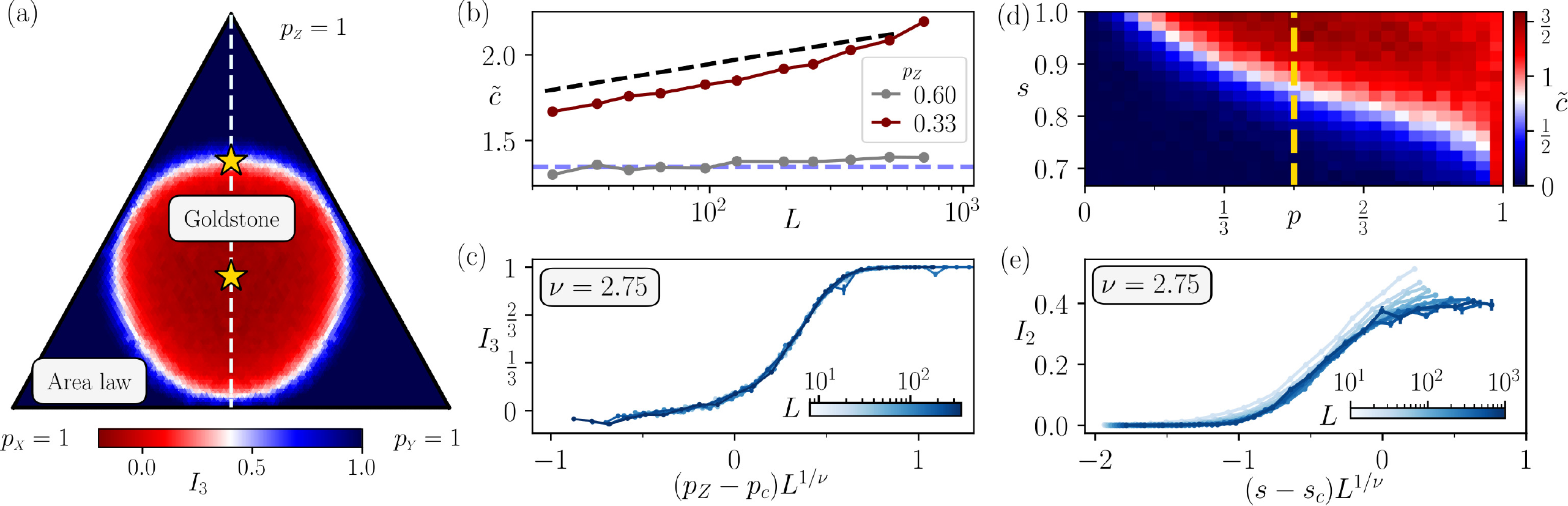}
    \caption{
    \textbf{Measurement phases in non-Gaussian, non-orientable circuits: }
    (a-c) Goldstone phase in the range-2 measurement-only circuit where Pauli operators $A_l B_{l+1}$ are measured with probability $p_A p_B$ with $A,B \in \{ X, Y, Z \}$.
    (a) Phase diagram of the tripartite mutual information $I_3$ showing a Goldstone phase (red) and area-law phase (blue).
    The white dashed line marks $p_X = p_Y$, along which we mark the upper critical point and the center of the Goldstone phase.
    (b) The prefactor $\tilde c$ of the logarithmic growth of the entanglement entropy.
    In the Goldstone phase (maroon) we observe a logarithmic correction with magnitude comparable to the universal $3\log(2)/(2\pi^2)$ expected for the CPLC Goldstone phase (black dashed line). 
    At the transition (gray) $\tilde{c}$ is independent of system size and consistent with $\tilde{c}\approx 1.33$ found along the critical line in CPLC.
    (c) The finite size scaling collapse of the tripartite mutual information $I_3$ confirms a critical exponent $\nu\approx 2.75$.
    (d-e) Goldstone phase in the interacting, non-orientable two-leg ladder with $p_{ZZ} = 1-s$, $p_Z = p_{XX} = p_{YY} = s(1-p)/3$, and $p_{XY} = p_{YX} = sp/2$.
    (d) Phase diagram in the $s$-$p$ plane showing a transition from a Goldstone phase at weak interactions to an area-law where four-fermion measurements dominate.
    (e) Scaling collapse of the mutual information $I_2$ along the dashed line in (d) marking $p=\tfrac12$.
    The scaling is consistent with a critical correlation length exponent $\nu\approx 2.75$, as expected from the Gaussian CPLC limit.
    }
    \label{fig:nonorientable_nonGaussian}
\end{figure*}

\emph{Dynamical purification:} A further signature of the Goldstone phase is logarithmic corrections in the mutual information between the two temporal boundaries, i.e., in the spanning number $n_s$~\cite{Nahum_2013}. 
At a conventional critical point with dynamical critical exponent $z=1$ and for fixed circuit aspect ratio $T=L$, $n_s$ is independent of $L$. 
This is confirmed at the critical points $p_{c,1/2}$ and shown in Fig.~\ref{fig:two_leg_ladder_results}c.
In the Goldstone phase, however, a logarithmic correction $n_s \propto \log(L)$ appears. 
This yields logarithmic corrections to the dynamical purification in the circuit when starting from a mixed initial state~\cite{Loio_2023, Fava_2023}.
As a result, there is an anomalous slowing of the purification, which is shown in Appendix~\ref{app:purification}.
We find a prefactor of the logarithmic correction of $0.19$, whereas $\tfrac{1}{2\pi}\approx0.159$ is expected from the renormalized spin-stiffness in CPLC~\cite{Nahum_2013}.

\emph{Bulk correlations:} Finally, we examine correlations in the spacetime bulk of the circuit, which are expressed via the two-leg watermelon correlator $G_2(L/2)$. 
At the critical points $p_{c,1/2}$ in the circuit, $G_2(L/2)$ shows slow power-law decay as $L^{-2x_2}$ with exponent $x_2 = 0.095$ consistent with the behavior in loop models with crossings~\cite{Nahum_2013}, see Fig.~\ref{fig:two_leg_ladder_results}d.
In the Goldstone phase, these correlations fall off even slower, decaying as $G_2(L/2) \propto \log(L/L_0)^{-\alpha_2}$ with a fitted exponent $\alpha_2 = 2.38 \pm 0.2$. 
Previous work found $\alpha_2 \approx 1.9$~\cite{Nahum_2013} and the slow decay limits the accuracy of the estimate.


\subsection{Non-Orientable Non-Gaussian Circuits} \label{sec:non_orient_non_gauss}
Here we examine the generality of the CPLC Goldstone phase in non-orientable quantum circuits by adding non-Gaussian measurements. 
We consider two cases: 
(i) adding orientable non-Gaussian measurements to a non-orientable Gaussian circuit, and
(ii) adding non-orientable non-Gaussian, e.g., parity breaking, measurements to an orientable Gaussian circuit. 
Orientability is then only broken by the non-Gaussian terms. 
Both cases yield an extended Goldstone phase with a universal $\log$-correction to the entanglement entropy. 
Since the loop model in the non-Gaussian circuits emerges only on a coarse grained scale, our main observables are multipartite mutual information measures for Clifford circuits, i.e., the bipartite and tripartite mutual information $I_2, I_3$. 

\subsubsection{Perturbing a CPLC with non-Gaussian measurements}

In order to confirm the expected irrelevance of non-Gaussian measurements in both the Goldstone and the area law phases, see Sec.~\ref{sec:beyond_ff}, we examine measurements of the set $\mathcal{O}_3\cup\{Z_l, Z_{l}Z_{l+1}\}$. 
This adds $ZZ$-measurements to the non-orientable setting of Sec.~\ref{sec:non_orient_Gauss}.
Note that $\mathcal{O}_3\cup\{ Z_{l}Z_{l+1}\}$ would be an orientable circuit. 
We set the measurement probabilities to $p_{ZZ} = (1-s)$, $p_{Z} = p_{XX} = p_{YY} = s(1-p)/3$ and $p_{XY}=p_{YX}=sp/2$, such that the circuit is described by parameters $p,s \in [0,1]$.
The Gaussian limit ($s=1$) is in a Goldstone phase for $0.22\lesssim p<1$. 
For $s<1$ the Goldstone phase remains robust against $ZZ$-measurements.
Eventually, non-Gaussian measurements become sufficiently frequent to drive the system into an area law phase. 
A scaling collapse of the mutual information at the transition confirms the correlation length exponent $\nu\approx 2.75$ as in CPLC (see Fig.~\ref{fig:nonorientable_nonGaussian}e). 
Though they may stabilize an area law phase at strong coupling, the non-Gaussian measurements are irrelevant at the CPLC fixed points, leaving the CPLC universality class unaltered.

\subsubsection{Inducing CPLC by non-Gaussian measurements} \label{sss:interaction_induced_gs}
Here we examine the emergence of CPLC, including the Goldstone phase, when adding non-orientable, non-Gaussian measurements to an orientable Gaussian circuit. 
In particular, we consider Pauli-measurements of two qubit operators $A_lB_{l+1}$ with $A,B\in\{X,Y,Z\}$, which form a range-2 measurement-only Clifford circuit~\footnote{The phase diagram of this model was previously obtained in Ref.~\cite{Ippoliti_2021}. However, no logarithmic correction was observed for the considered system sizes.}. 
The emergent CPLC in this model further confirms that short-ranged measurement-only Clifford circuits yield emergent loop model behavior, even when including non-Gaussian, parity violating measurements. 
Each operator is measured with probability $p_{AB}=p_ap_b$, i.e., $p_{XY}=p_xp_y$, with $p_x+ p_y + p_z = 1$. 
A global rotation $U_{A}\equiv\exp\left(i\tfrac{\pi}{4}\sum_l A_{l}\right)$ with $A\in\{X,Y,Z\}$ maps between parameters $p_x\overset{U_Z}{\leftrightarrow} p_y\overset{U_X}{\leftrightarrow} p_z\overset{U_Y}{\leftrightarrow} p_x$, yielding a highly symmetric circuit and phase diagram, which is shown in  Fig.~\ref{fig:nonorientable_nonGaussian}a.

By virtue of the symmetry above, we can focus on the parameter regime $p_x,p_y\gg p_z$, which corresponds to perturbing the orientable Gaussian circuit $\mathcal{O}_3=\{XX, YY, XY, YX\}$ with parity breaking non-Gaussian operators $\{XZ, YZ, ZX, ZY\}$.
The parity breaking operators also break the orientability of the worldlines, as discussed in Sec.~\ref{ss:parity_breaking}. 
Along the orientable boundary $p_z = 0$, we recover the bipartite two-leg ladder, which lies in an area-law phase for all $p_x + p_y = 1$.
As in conventional CPLC, the short-loop area-law phase is robust against a small probability of orientability (parity) violating measurements.
Then at small but non-zero $p_z^{(c)} \approx 0.04$, there is a transition into a CPLC Goldstone phase.
The Goldstone phase extends over a large fraction of the parameter range and generally appears when orientability breaking by parity violating measurements becomes sufficiently strong. 
The characteristic $~(\log(L))^2$-growth of the entanglement entropy is shown in Fig.~\ref{fig:nonorientable_nonGaussian}b. 
The critical line separating the Goldstone from the area law phase features the correlation length exponent $\nu=2.75$ expected of the CPLC universality class, confirmed in Fig.~\ref{fig:nonorientable_nonGaussian}c.

We note that taking the above example and replacing $ZX, XZ, ZY, YZ$-measurements by either $X$- or $Y$-measurements alone \emph{neither} induces a non-orientable circuit \emph{nor} a Goldstone phase. 
The effective swaps emerging from $X$ or $Y$ measurements preserve orientability when added to the set $\mathcal{O}_3$, since products of subsequent measurements, e.g., $X_l X_{l'}$, are obtained from a sequence of measurements from $\mathcal{O}_3$. 
The same applies to $4$-qubit operators acting on four consecutive qubits, e.g., $XZXZ, ZXXZ$. 
The simplest non-trivial, orientability breaking terms arising from the random two-qubit measurements are the non-Gaussian operators $Z_l X_{l+1} Y_{l+2} \sim Z_l Y_{l+1} \times Z_{l+1} Y_{l+2}$.
We demonstrate in Appendix~\ref{app:interaction_induced_gs} that including the latter instead of parity breaking operators indeed yields a CPLC Goldstone phase.

\section{Space-Time Duality}\label{sec:duality}
Each $(1+1)$-dimensional quantum circuit possesses a space-time dual circuit, in which the measurement- and swap-vertices $\calP, \calR$ are rotated around $\tfrac{\pi}{2}$ in space-time, i.e., for which space and time are exchanged. 
The loop model of the space-time dual circuit displays the same bulk behavior, while its spatial boundary correlations, such as the mutual information, are exchanged with the temporal boundary correlations $\sim\abs{ \langle \gamma_1(t_2) \gamma_1(t_2) \rangle }$. 
Depending on the particular loop model, it may be advantageous to implement the space-time dual instead of the original circuit, e.g., in order to overcome the postselection problem in non-Clifford circuits~\cite{Ippoliti_2021_duality, Lu_2021_spacetime, Ippoliti_2022_spacetime, ippoliti_google_2023}. 
Alternatively, one may implement ``isotropic'' or space-time self-dual circuits for which the space-time dual and the original circuit are identical. 
This yields an additional symmetry, which may grant access to an exact analytical solution of the circuit evolution~\cite{Reid_2021, Prosen_2021, Jonay_2021_triunitary, Claeys_2022_emergentquantum, Kasim_2023, Claeys_2022, Foligno_2022, Stephen_2022, Rampp_2022, Claeys_2023, Sommers_2023}.

Consider for instance a circuit with nearest neighbor Majorana operations, involving only 4-leg Majorana vertices. 
Then the space-time rotation (i) maps each swap $\calR_{l,l+1} \to \calR_{l,l+1}$ onto itself, i.e., swaps are self-dual and (ii) exchanges the identity and the projection $\mathds{1}_{l,l+1}\leftrightarrow \calP_{l,l+1}$. 
Therefore the circuit is space-time self-dual when both $\mathds{1}, \calP$  appear with equal probability. 

\begin{figure}[!t]
    \centering
    \includegraphics[width=\columnwidth]{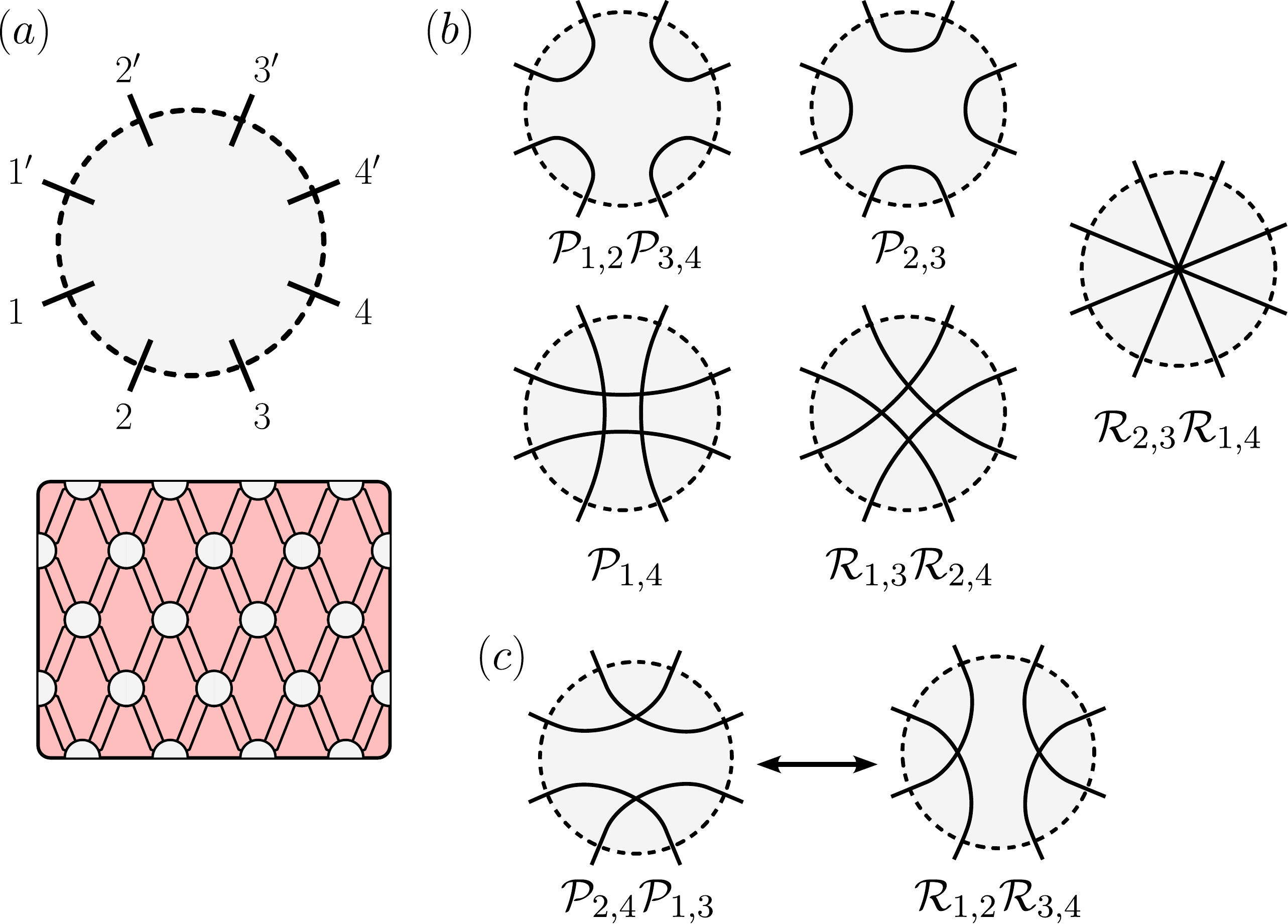}
    \caption{
    \textbf{Space-time duality in a brickwork circuit.}
    (a) 
    Alternating layers of two-qubit operations give a tilted square lattice where each edge carries two worldlines.
    Each ``brick'' in the circuit consists of a generic 8-leg vertex in which the four incoming (unprimed) and outgoing (primed) worldlines must be paired amongst one another.
    Exchanging space and time gives a $\pi/2$ rotation of such an 8-leg vertex configuration.
    (b)
    The five 8-leg vertex configurations which are self-dual under the space-time rotation, along with the corresponding operation.
    If instead a $\pi/4$ rotation is made, then two pairs rotate into one another while the 4-line crossing $\calR_{2,3}\calR_{1,4}$ is left invariant.
    (c)
    A pair of vertex configurations which transform into one another under the space-time rotation.
    For this example, parity checks are transformed into swaps, giving an intuitive picture for how measurements act to mediate effective swaps.
    }
    \label{fig:spacetime_duality_figure}
\end{figure}

Allowing nearest neighbor qubit operations composed of four Majorana fermions requires eight-leg Majorana vertices in space-time. 
Let us consider a brickwall circuit consisting of alternating layers of ``bricks'' acting on four neighboring fermions, as depicted in Fig.~\ref{fig:spacetime_duality_figure}a. 
Each brick is an 8-leg vertex in the loop model, with four incoming and four outgoing worldlines, and the set of possible bricks consists of all ways to pair the 8 worldlines with one another.
Under space-time rotation, 8-leg vertices transform into one another by permuting the legs (e.g. $1 \rightarrow 2', 2\rightarrow 1',3 \rightarrow 1, \dots$ in Fig.~\ref{fig:spacetime_duality_figure}a). 
Amongst all 8-leg vertices with unambiguous worldline representation, only five are self-dual, see Fig.~\ref{fig:spacetime_duality_figure}b.

Let us denote $\mathcal{V}_8$ as the set of all possible configurations of worldline pairings in an 8-leg vertex.
The space-time rotation induces a permutation $\sigma$ on $\mathcal{V}_8$ such that any vertex $v \in \mathcal{V}_8$ is transformed into $\sigma(v) \in \mathcal{V}_8$. 
Self-dual vertices have $\sigma(v)=v$, while vertices $v_1$ invariant under a $\pi$ rotation but not under a $\pi/2$ rotation have a dual vertex $v_2$ such that $v_1 \xleftrightarrow{\sigma} v_2$.
The remaining vertices can then be separated into groups of four such that $v_1 \xrightarrow{\sigma} v_2 \xrightarrow{\sigma} v_3 \xrightarrow{\sigma} v_4 \xrightarrow{\sigma} v_1$, since $\sigma^4=\mathds{1}$. 
In a space-time self-dual circuit, every vertex $v$ and its dual $\sigma(v)$ occur with the same probability $P$, i.e., $P(\sigma(v))=P(v)$ for all $v\in\mathcal{V}_8$. 
This gives rise to an abundance of Majorana brickwall circuits, which feature both spacetime duality and an exact worldline description. 
\begin{figure*}[!ht]
    \centering
    \includegraphics[width=\textwidth]{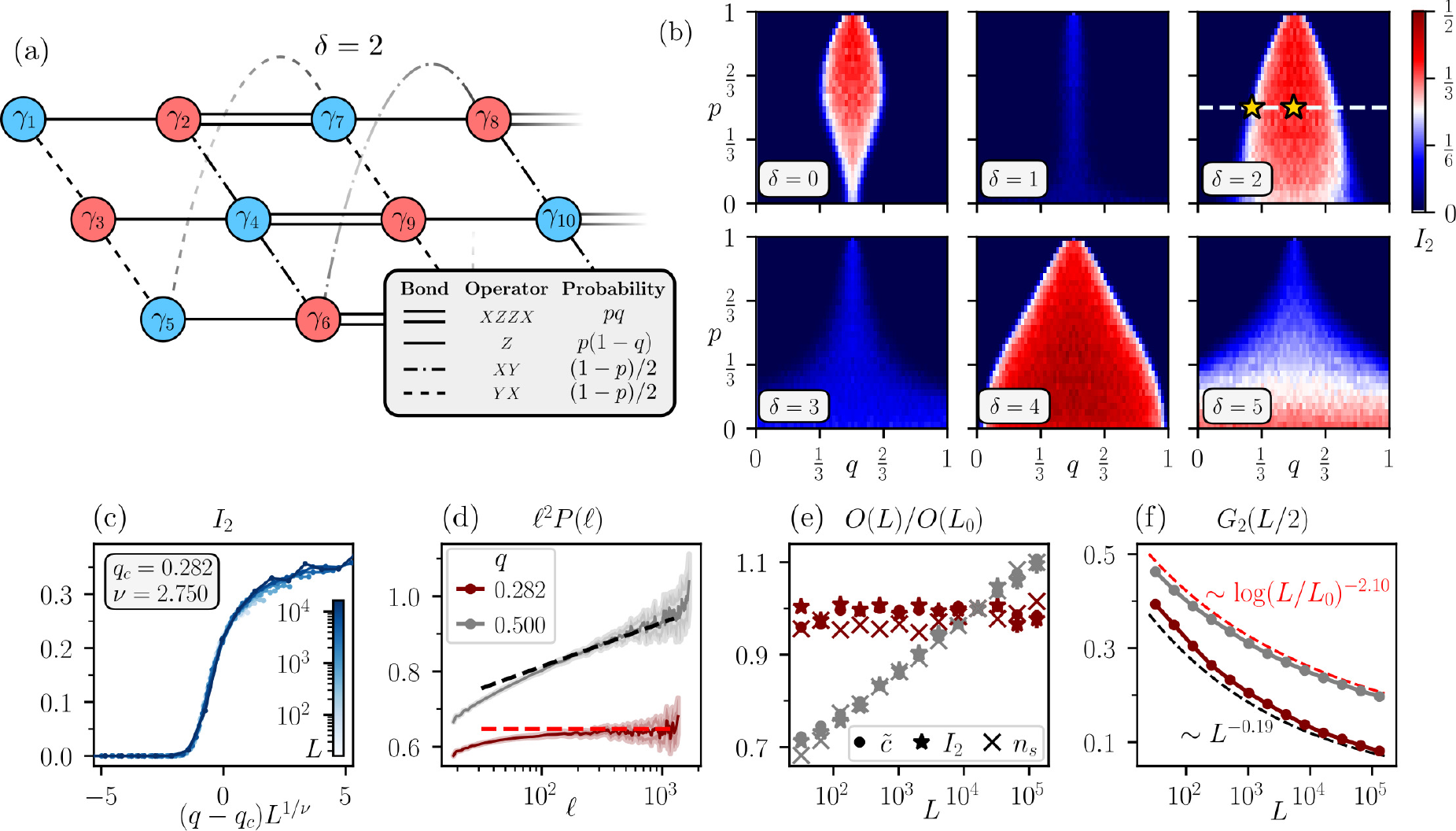}
    \caption{
    \textbf{Loop model for the XZZX-code:}
    (a) Majorana lattice where each bond corresponds to an allowed parity-check on the 3-chain square lattice. 
    The measurement probabilities for the different types of bonds and the corresponding Pauli operator on a single qubit chain are shown in the legend and parameterized by $p,q \in [0,1]$.
    Transverse boundary-conditions defined by the twist $\delta$ correspond to connecting the first and third chains with bonds $\gamma_{(n,3)} \leftrightarrow \gamma_{(n+\delta, 1)}$ (here $\delta = 2$).
    (b) The phase diagram in the $p$---$q$ plane depicted via the mutual information $I_2$ at $L=12288$ shown for different boundary twists $\delta$.
    For even twists, the lattice is non-bipartite and we observe an extended Goldstone phase (red).
    For odd twists, the lattice remains bipartite and the worldlines are orientable such that no Goldstone phase is observed.
    Here the mutual information is finite only when crossing $q=\tfrac12$ or $p=0$, corresponding to percolation criticality.
    (c-f) Entanglement and loop statistics with $\delta = 2$ along the dashed $p=\tfrac12$ line in (b).
    (c) Scaling collapse of the two-interval mutual information $I_2$ near transition identifies the critical point $q_c \approx 0.282$ and correlation length exponent $\nu=2.75$.
    (d) The steady-state stabilizer length distribution $P(\ell)$ reveals a logarithmic correction in the Goldstone phase compared to the conventional $P(\ell) \propto \ell^{-2}$ scaling at the transition.
    Dashed lines correspond to the universal value expected from CPLC in the Goldstone phase (black) and at the transition (red) \cite{Nahum_2013}.
    (e) The system-size dependence of the mutual information $I_2$, fitted log-law coefficient $\tilde{c}$, and spanning number $n_s$, normalized against their magnitude at $L_0=2^{14}$, reveals logarithmic corrections in the Goldstone phase that are absent at the transition.
    (f) The two-leg watermelon correlator $G_2(L/2)$ exhibits power-law decay at the transition, with exponent $x_2 \approx 0.095$ consistent with that found in CPLC.
    Correlations decay more slowly in the Goldstone phase, falling off as $\log(L/L_0)^{-\alpha_2}$ with $\alpha_2 \approx 2.1$.
    }
    \label{fig:three_leg_results}
\end{figure*}

A peculiar scenario arises when $v$ and $\sigma(v)$ correspond to different types of operations, i.e., when $v$ results from a non-unitary operation while $\sigma(v)$ is unitary. 
This gives rise to the possibility to implement an entirely unitary circuit, which realizes the space-time dual of a non-unitary circuit including measurements. 
This approach has been demonstrated for Clifford circuits in order to overcome the post-processing~\cite{Noel_2022} or post-selection~\cite{minnich_postselect} problem of non-deterministic measurements~\cite{Lu_2021_spacetime, ippoliti_google_2023}. 

In the loop model framework, a large class of both orientable and non-orientable circuits can be simulated by purely unitary evolution. 
However, since the space-time dual of the identity is a projective parity measurement, only ``dense'' non-unitary circuits, where each brick realizes either a swap or a measurement but never the identity, have a unitary space-time dual. 
This poses a restriction on the corresponding loop model. 
For instance the CPLC realized by purely unitary operations lives at the border of the CPLC phase diagram~\cite{Nahum_2013} and cannot stabilize a Goldstone phase. 
Instead, it realizes a gapped area law phase in both the temporal and the spatial direction, with a parametrically large correlation length. 
This can be rationalized by the fact that in 2D a bulk phase which breaks a continuous symmetry is only allowed under a non-unitary evolution.

Phase transitions which continuously break a discrete symmetry do not have this constraint in two dimensions. 
This grants access to the physics of coupled Potts models, e.g., to the Ising and percolation phase transitions, are observed in orientable circuits also via a purely unitary evolution.

\section{Loop models in different geometries}\label{ss:three_leg_ladders}

We now turn our attention to Gaussian measurement-only circuits for which we lift the constraint that measurements be limited to nearest and next-nearest neighbors.
This allows us to manipulate the geometry of the underlying Majorana lattice in order to probe the presence or absence of a Goldstone phase as a function of the circuit's orientability.
The starting point here is measurements of the operators $Z, XY, YX$ and the $XZZX$ stabilizer.
This relates the setup to an important class of quantum error correcting codes~\cite{Bonilla_XZZX} and to topological phase transitions: for suitable measurement rates, the circuit hosts both a symmetry protected topological order and a bulk symmetry breaking order~\cite{Klocke_2022, morral_2023}.

In the presence of $XZZX$-stabilizer measurements, it is convenient to arrange the Majorana fermions on a three-leg ladder geometry as shown in Fig.~\ref{fig:three_leg_results}a.
The $Z$- and $XZZX$-measurements implement three copies of the repetition code~\cite{Klocke_2022} or alternatively a measurement-only version of the fermionic 3-chain obtained by interleaving three Kitaev chains~\cite{Verresen_2017}.  
The chains are coupled via $XY$ and $YX$ Pauli measurements connecting $\gamma_{l}$ and $\gamma_{l+2}$. 
This implements vertical bonds between chains $1\leftrightarrow2$ and $2\leftrightarrow3$, while it creates a relative offset at the boundary, coupling mode $n$ on chain $3$ and mode $n+\delta$ with $\delta=2$ on chain $1$. 
This link, i.e., the offset $\delta=2$, makes the circuit non-bipartite and its loop model non-orientable. 
Conversely, an offset $\delta=1$ instead of $\delta=2$ would make the graph bipartite and restore orientability. 

This argument can be extended to generic $m$-leg ladders with rectangular inter-chain bonds between chains $k=1,\dots m-1$. 
Making such a graph bipartite requires N\'{e}el ordering of the worldline orientations. 
This is only compatible with an offset $\delta$ between the $m$-th and the $1$st chain if $\delta+m$ is even. 
Otherwise the ladder is \emph{not} bipartite and orientability is broken.

We numerically confirm the presence (absence) of a Goldstone phase for non-orientable (orientable) ladder geometries in Fig.~\ref{fig:three_leg_results}b.
It shows the entanglement in the circuit for different offsets $\delta$ and measurement probabilities.
The latter are parameterized by $p, q \in [0,1]$ as shown in  Fig.~\ref{fig:three_leg_results}a.
Here $\tfrac12(1-p)$ is the probability for measuring one of the two inter-chain bonds and $q$ sets the dimerization of measurement probabilities along each chain, i.e., $p=1,q=\tfrac12$ is the percolation critical-point of the $XZZX$-code~\cite{Klocke_2022}. 
When $\delta$ is odd, only the lines $q=\tfrac12$ and $p=0$ show increased mutual information, with $p=0$ corresponding exactly to $\delta$ copies of critical percolation.
For $\delta$ even, we instead observe an extended Goldstone phase. 
When increasing $\delta$, the parity measurements across the boundary of the ladder become longer in range and facilitate the onset of the Goldstone phase already at small measurement probabilities.

We show in Fig.~\ref{fig:three_leg_results}b that turning on finite inter-chain coupling, i.e., breaking the orientability of the ladder, causes the percolation critical point to broaden into a critical fan, which is symmetric about $q=\tfrac12$.
Upon examination, we confirm that the entangled phase inside the fan is the Goldstone phase of CPLC. In the remainder of this section we verify the universal behavior of CPLC for the particular case of $\delta = 2$.

\emph{Correlation length critical exponent:} The scaling collapse of the mutual information $I_2$ along the dashed line at $p=\tfrac12$ in Fig.~\ref{fig:three_leg_results}c identifies the critical points $q_c = 0.282, 0.718$ (symmetric around $q=0.5$) and the CPLC critical exponent $\nu=2.75$.

\emph{Loop lengths:} We define the length $\ell$ of a loop with respect to the indexing of the fermions on the lattice.
The transition, $q=q_c$, features conventional critical scaling $\ell^2P(\ell) \to \alpha$, while the Goldstone phase exhibits logarithmic corrections $\ell^2P(\ell) \to \beta \log(\ell)$ with both $\alpha$ and $\beta$ being consistent with the CPLC values, see Fig.~\ref{fig:three_leg_results}d.

\emph{Entanglement growth:} The loop distributions $P(\ell)$ readily imply the behavior for the entanglement entropy $S_A$ and the mutual information $I_2(A,B)$ of contiguous subsystems of size $|A|, |B|$. 
We again take the ansatz $S_A(L) = \frac{\tilde{c}(L)}{3} \log_2\left[\frac{L}{\pi}\sin\left(\frac{\pi |A|}{L}\right)\right] + \text{const}$. 
At the critical points, $\tilde{c}, I_2$ are constant in system size, see Fig.~\ref{fig:three_leg_results}e, while a logarithmic correction is found in the Goldstone phase, $\tilde{c}, I_2 \propto \log(L)$. 
The numerical prefactors follow from $P(\ell)$ and are thus consistent with CPLC.

\emph{Purification:} The Goldstone phase exhibits logarithmic corrections to purification, witnessed via the spanning number.
In Fig.~\ref{fig:three_leg_results}e we show the spanning number $n_s$ at fixed circuit aspect ratio $T = L$, which remains constant at $q_c$ but grows logarithmically with $L$ in the Goldstone phase.
The prefactor of the logarithmic growth is $0.14$, comparable to $\tfrac{1}{2\pi} \approx 0.16$ expected from CPLC.
See App.~\ref{app:purification} for additional details on the dynamical purification and the spanning number.

\emph{Bulk critical exponents:} Using the two-leg watermelon correlator $G_2(L/2)$, we extract bulk critical exponents at the critical point and in the Goldstone phase (see Fig.~\ref{fig:three_leg_results}f).
At the transition, the results are consistent with $G_2(L/2)\sim L^{x_2}$, with exponent $x_2 \approx 0.095$.
By contrast, we observe a logarithmic slowing down $G_2 \sim \log(L/L_0)^{-\alpha_2}$ with exponent $\alpha_2 \approx 2.1$ in the Goldstone phase.
Both exponents are consistent with the results in the two-leg ladder and CPLC.


\section{Conclusion}\label{sec:discussion}

Two-dimensional loop models represent a paradigm for a large class of solvable models in quantum and classical statistical mechanics. 
Using analytical arguments, confirmed by numerical simulations, we have strengthened the link between measurement-only Clifford circuits in (1+1) dimensions and 2D loop models. 
Our work demonstrates that loop models provide a general framework which covers a large class of measurement-only circuits and measurement-induced phase transitions, including previously studied models as well as new types of setups considered here.
Mapping circuits to loop models provides a means to classify the measurement-only evolution based on the symmetry and the topology of the corresponding loop model, as well as to analyze the universal long-wavelength behavior at measurement-induced phase transitions. 

The loop model approach outlined in our work, in particular the aspect of worldline orientability, provides a blueprint to study the rich phenomenology of 2D loop models with (1+1)-dimensional Clifford circuits. 
Promising routes may be the realization of exotic entangled states, such as topological phases or larger classes of non-unitary conformal field theories, which both are known to admit mappings to orientable loop models. 
The emergence of an entanglement transition with Ising universality at the tricritical point of the 1-state Potts model stands as a particularly clear example of the diverse set of states and critical dynamics realizable by employing the loop model framework for quantum circuits.

Further exploring the link to loop models opens up a new perspective for general monitored quantum circuits. The question arises naturally, whether it is possible to extend the loop model framework to include  general unitary operations beyond discrete swap gates or non-Clifford measurements. For the former, the particular case of random Majorana gates has been shown to yield a similar nonlinear sigma model as the one that is obtained for CPLC~\cite{Fava_2023,Jian_2023}. Including non-Clifford measurements may similarly give rise to new braiding or fusion rules, which go beyond classical loop models. In both cases, orientability may only emerge on certain fine-tuned parameter values, while new symmetries may emerge, which enlighten the connection of monitored circuits to quantum loop model approaches~\cite{Trebst_2007, Aasen_2020}.

Based on the current work, an exciting future direction is the study of measurement-only circuits in higher dimensions and for different geometries. For instance in two-dimensional circuits, the loop model and in particular its symmetries may be enriched by tunable geometries. An example of the latter is found in the 2D measurement-only Kitaev spin liquid~\cite{Zhu_2023,Lavasani_2022}, which gives rise to a tripartite Majorana graph. Adding higher order Majorana measurements breaks this symmetry and makes the graph non-orientable, which leads to a peculiarly entangled Majorana liquid~\cite{Zhu_2023}. Exploring this scenario and generalizations with the loop model approach will help to establish a general classification of measurement-only circuits in higher dimensions beyond the currently established paradigms.

\begin{acknowledgements}
We thank Joel Moore for fruitful discussions.
KK was supported by an NSF Graduate Fellowship under Grant No. DGE 2146752 and by the U.S. Department of Energy, Office of Science, National Quantum Information Science Research Centers, Quantum Science Center. 
MB acknowledges support from the Deutsche Forschungsgemeinschaft (DFG, German Research Foundation) under Germany’s Excellence Strategy Cluster of Excellence Matter and Light for Quantum Computing (ML4Q) EXC 2004/1 390534769, and by the DFG Collaborative Research Center (CRC) 183 Project No. 277101999 -project B02.
\end{acknowledgements}

\nocite{apsrev42Control}
\bibliographystyle{apsrev4-2}
\bibliography{main}


\begin{appendix}

\section{Numerical Method for Majorana Circuits}\label{app:computation_details}

Whereas a stabilizer state in a generic Clifford circuit requires an $O(L^2)$ tableau representation, the current state of a Gaussian Majorana circuit is represented by the $O(L)$ pairings corresponding to the endpoints of worldlines.
Moreover, while the computational cost of a projective measurement generally scales as $O(L^3)$ for stabilizer states, measurements in a loop model require only $O(1)$ operations to rearrange the pairings.
These computational benefits are further enhanced by efficient schemes for generating deeper circuits via shuffling and concatenating shallower circuits \cite{Nahum_2013}.
By generating a ``pool'' of $N$ shallow circuits (e.g. a single layer), repeated shuffling and concatenation of the elements doubles the circuit depth with every round of concatenation.
Then reaching a circuit depth $T$ for $L$ qubits in a pool of size $N$ requires computational cost that scales as $\mathcal{O}(NL\log_2(T))$.
As a result, one can efficiently simulate system sizes several orders of magnitude greater than is possible with stabilizer states.
This is crucial for accurately identifying logarithmic scaling corrections which are otherwise challenging to distinguish from anomalous power-law scaling.

\subsection{Entanglement statistics from the boundary-loop distribution}

Given a circuit trajectory with periodic boundary conditions, we might ask for an entanglement measure averaged over all translations of the system in order to extract maximal information out of the state.
In this case, one can equivalently take the distribution $P(\ell)$ of open loops with both ends on the final $t=T$ temporal boundary, reducing the number of computations which need to be performed in each trajectory.
Here we give a short review of how this is implemented.

Let $A$ and $B$ be two \emph{disjoint} subsystems for which we would like to compute the ensemble averaged mutual information $\llangle I_2(A,B) \rrangle$.
This can be done via the sum
\be
    I_2(A,B) = \frac12 \sum_{x \in A} \sum_{y \in B} P(\ell = \textrm{dist}(x,y)),
\ee
where the distance function respects the PBC such that
\[
    \textrm{dist}(x,y) = \textrm{min}(\abs{x-y}, N - \abs{x - y}).
\]
and the factor of $1/2$ accounts for the directionality of the distance function.
Without loss of generality, let $A = [1, \abs{A}]$ and $B = [\abs{A} + \Delta + 1, \abs{A} + \Delta + \abs{B}]$ with $\abs{A}, \abs{B} \leq N/2$ and $\Delta \leq N - \abs{A} - \abs{B}$ so that the intervals remain disjoint.
We then always have $y > x$ for $y \in B$ and $x \in A$, allowing us to write the distance function instead as
\[
    \textrm{dist}(x,y | y > x) = \begin{cases} y - x & y - x \leq N/2, \\ N - (y-x) & \textrm{else.}\end{cases}
\]

In practice, the double sum over coordinates is inefficient when going to very large system sizes.
This can be simplified by instead expressing the mutual information as 
\[
    I_2(A,B) = \sum_{x\in A, y \in B} P(\ell = \textrm{dist}(x,y)) = \sum_{\ell = 1}^{N/2} P(\ell) w_{A,B}(\ell)
\]
where the weight function $w_{A,B}(\ell)$ depends on the partitions $A,B$ only via $\abs{A}$, $\abs{B}$, and $\Delta$ and can be written explicitly as 
\[
    w_{A,B}(\ell) = \frac12 \sum_{x\in A, y\in B} \delta_{\ell, \textrm{dist}(x,y)}.
\]
We now give explicit derivation of $w(\ell)$ for two cases (i) the computation of $I_2$ and (ii) the computation of $S_\ell$.

\subsubsection{Weight function for two-interval mutual information}

Recall that throughout the main text we computed the mutual information $I_2$ between regions $A = [1, N/8]$, $B = [1 + N/4, 3N/8]$ such that $\abs{A} = \abs{B} = \Delta = N/8$.
From this choice, the distance function is always given simply by $y - x$.
Letting $y = z + 2\abs{A}$ we can evaluate the weight function
\be
    \begin{aligned}
        w(\ell) &= \frac12 \sum_{x = 1}^{\abs{A}} \sum_{z = 1}^{\abs{A}} \delta_{\ell, 2\abs{A} + z - x}, \\
        &= \frac12 \left(\abs{A} - \abs{2\abs{A} - \ell}\right) \Theta(\abs{A} - \abs{2\abs{A} - \ell})
    \end{aligned}
\ee
where $\Theta(\cdot)$ is the step function.
This gives exactly the number of positions $x \in A$ such that a loop of length $\ell$ terminates at some $y = x + \ell \in B$.
Then to compute the ensemble averaged entanglement entropy we can simply compute the weighted sum
\[
    \llangle I_2(A,B) \rrangle = \sum_\ell P(\ell) w(\ell)
\]
using the loop-length distribution obtained from averaging over many trajectories.

\subsubsection{Weight function for the subsystem entanglement}

While a single computation such as $I_2$ is generally inexpensive, it becomes quite costly if one wants the subsystem entanglement entropy $S_\ell$ for all allowed subsystem sizes $\ell$.
As such it is useful to seek an efficiently computable expression for $w(\ell)$.
Let us now take subsystem $A = [1,\abs{A}]$ and its complement $\overline{A} = [\abs{A} + 1, N]$ for $\abs{A} \leq N/2$.
It is no longer generically true that $y - x \leq N/2$ for $y \in \bar{A}$ and $x \in A$ and so some additional care is required.
For fixed $x \in A$, the distance function switches behavior at $y^* = \frac{N}{2} - \abs{A} + x \geq 1$.
We can then split the sum over $y$ and evaluate $w(\ell)$ as follows
\be
    \begin{aligned}
        w(\ell) &= \frac12 \sum_{x = 1}^{\abs{A}} \sum_{z = 1}^{N - \abs{A}} \delta_{\ell, \textrm{dist}(x, \abs{A} + z)} \\ 
        &= \frac12 \sum_{x=1}^{\abs{A}} \left[ \sum_{z=1}^{z^*} \delta_{\ell, \abs{A} + z - x} + \sum_{z = z^* + 1}^{N - \abs{A}} \delta_{\ell, N - \abs{A} - z + x} \right]\\
        &= \begin{cases}
            \textrm{min}(\ell, \abs{A}) & 1 \leq \ell < N/2 \\
            N/2 & \ell = N/2.
        \end{cases}
    \end{aligned}
\ee
Then the ensemble averaged entanglement entropy for a contiguous subsystem of $\abs{A} = \ell$ Majoranas is 
\[
    \llangle S_\ell \rrangle = \frac{N}{2} P(N/2)  + \sum_{\ell' = 1}^{\ell - 1} \ell' P(\ell') + \sum_{\ell' = \ell}^{N/2 - 1} \ell P(\ell').
\]

While this has eliminated one sum (and many calculations from each trajectory), it is still cumbersome to do $O(N^2)$ calculations to compute $\llangle S_\ell \rrangle$ for all $\ell$.
This can be made even simpler by observing a recursive structure with respect to $\ell$.
In particular we have 
\[
    \llangle S_{\ell + 1} \rrangle = \llangle S_{\ell} \rrangle + \sum_{\ell' = \ell + 1}^{N/2 - 1} P(\ell').
\]
Now let us define a slightly modified cumulative density function for the loop length distribution,
\[
    F(\ell) \equiv \sum_{\ell' = \ell}^{N/2 - 1} P(\ell'),
\]
such that $F(1) = 1 - P(N/2)$.
Then we can write the following recursive relations
\[
    \begin{aligned}
        F(\ell + 1) &= F(\ell) - P(\ell), \\
        \llangle S_{\ell + 1} \rrangle &= \llangle S_{\ell} \rrangle + F(\ell + 1).
    \end{aligned}
\]
Equipped with the above, the full set of $\llangle S_\ell \rrangle$ can be computed from the loop length distribution $P(\ell)$ in $\mathcal{O}(N)$ time without any computation during individual trajectories.

\subsubsection{Watermelon Correlators}

In principle, the watermelon correlator $G_2(x,y)$ can be extracted from the ensemble averaged loop length distribution $P(\ell)$.
However in practice this is quite cumbersome as we outline below.
Given a circuit of depth $T$, let $P(\ell)$ be the distribution of boundary loops on the final time boundary, assuming that we imposed fixed boundary conditions at $t=0$ (i.e. a pure initial state).
For $x=(l, T)$ and $y=(m,T)$, $G_2(x,y)$ gives the probability that $x$ and $y$ lie along the same loop in the \emph{bulk} of a loop model.
Taking two copies of the circuit, we can glue the open ends together in order to close all of the loops.
Then $G_2(x,y)$ is the probability that points $l$ and $m$ on the boundary of the circuit are connected by a loop when these two copies of the circuit are glued together.
A closed loop formed by the gluing consists of alternating arcs from the two copies, with the number of arcs in the path being between $2$ and $2N$.
The probability of $l$ and $m$ being on the same loop can be decomposed into a sum over the probability of connecting $\ell$ and $m$ with a path of fixed number of arcs,
\[
    G_2(l, m) = \sum_{i=1}^{N} \textrm{Prob}\left(l \leftrightarrow m \vert \, 2i \textrm{ arcs} \right) \equiv \sum_{i=1}^N \mu_i(l \leftrightarrow m)
\]
For example the first order term $\mu_i$ is simply given by $P(\abs{l-m})^2$.
Higher order terms can be further broken into a sum over the length of the two halves of the path (i.e. $l \rightarrow m$ and $m \rightarrow l$).
In general there are an exponential number of possible paths, making the calculation of $G_2(l,m)$ from $P(\ell)$ intractible.
Of course many such paths have negligible weight and could be excluded from the sum, but this remains impractical.
As such, all values of $G_2$ reported in this work are computed in \emph{each} trajectory rather than from the ensemble averaged length distribution $P(\ell)$.

\section{Dynamical Purification Timescale}\label{app:purification}

As we have noted throughout the main text, in the Goldstone phase there are logarithmic corrections to the dynamics which can be observed via the spanning number $n_s(t,L)$.
In Fig.~\ref{fig:two_leg_ladder_results}c and Fig.~\ref{fig:three_leg_results}e, we show for fixed circuit aspect ratio that the spanning number $n_s(t=L, L)$ grows logarithmically with the system size.
As noted in Ref.~\onlinecite{Nahum_2013}, for unit aspect ratio ($t=L$) the spanning number should be given exactly by the renormalized spin stiffness of the $O(n=1)$ loop model at RG scale $t=L$.
This result can be readily extended to generic aspect ratio.
At RG scale $\Lambda$, the renormalized spin-stiffness is given to leading order by $\tilde{K}(\Lambda) = \tfrac{1}{2\pi}\log\left(\Lambda / \Lambda_0\right)$.
As a result, we expect the spanning number in the Goldstone phase to evolve as 
\be
    n_s(t,L) \sim F\left(\frac{t}{L} \frac{2\pi}{\log(L/L_0)} \right)
    \label{eq:span_num_evo}
\ee
for some scaling function $F$.
In the circuit language, this corresponds to a slowing of purification such that the typical timescale over which a maximally mixed state is purified now depends on system size as $L\log(L)$ rather than $L$.

\begin{figure}[!ht]
    \centering
    \includegraphics[width=\columnwidth]{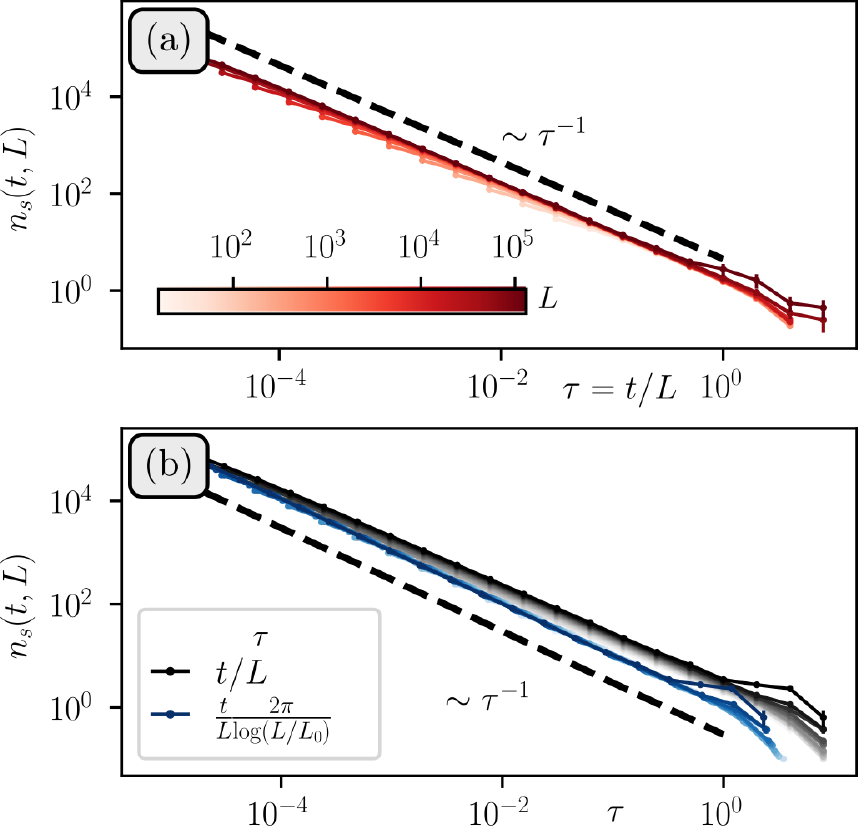}
    \caption{
    Evolution of the spanning number $n_s(t,L)$ in the three-leg ladder at the marked points in Fig.~\ref{fig:three_leg_results}b.
    (a) The transition at $p=\tfrac12$, $q=q_c = 0.282$ has conventional critical behavior with dynamical exponent $z=1$.
    This is reflected by the spanning number collapsing to a single curve parameterized by rescaled time $\tau = t / L$.
    While $n_s > 1$, purification gives $n_s \propto \tau^{-1}$, as shown by the black dashed line.
    (b) In the Goldstone phase at $p=q=\tfrac12$, logarithmic corrections modify the purification dynamics.
    Here we show $n_s(t,L)$ for two different rescaled times $\tau$.
    Conventional $z=1$ scaling with $\tau=t/L$ (gray) does not capture the logarithmic corrections.
    Accounting for the renormalized spin stiffness by taking $\tau = \tfrac{2\pi t}{L\log(L/L_0)}$ with $\log(L_0) = -5$ yields excellent data collapse.
    }
    \label{fig:spanning_number_evo}
\end{figure}

In Fig.~\ref{fig:spanning_number_evo} we show the time-evolution of the spanning number in the three-leg ladder at the two marked points from Fig.~\ref{fig:three_leg_results}b.
At the transition, $q=q_c$, we observe conventional critical purification with dynamical exponent $z=1$ such that $n_s(t,L) = F(\tau = t/L)$.
Equivalently, this can be considered as the $L_0 \rightarrow 0$ limit of Eq.~\eqref{eq:span_num_evo}.
In the Goldstone phase, at $q=\tfrac12$, the logarithmic corrections lead to finite $L_0$ being required for scaling collapse of the spanning number.
For this particular example we find $\log(L_0) = -5$, which implies that logarithmic corrections are non-negligible for $L\gtrsim 10^2$.

For small system sizes, these logarithmic corrections to purification can be difficult to unambiguously identify.
Indeed, a naive rescaling as $n_s(t,L) \sim F\left[tL^{-z^*}\right]$ in the regime $n_s \gtrsim 1$ would suggest an anomalous scaling exponent $z^*\approx 1.1$.

\section{Effective Transfer Matrix for Interacting Majorana Circuits}\label{app:effective_transfer_matrix}

In this section we derive effective transfer matrix description for the family of orientable, interacting Majorana circuits considered in Sec.~\ref{sec:orient_non_gauss}.

\subsection{Measurement-only $XXZ$-chain}

Let us consider the decoupled limit of the two-leg ladder ($XY$ and $YX$) perturbed by four-fermion ($ZZ$) measurements.
Moreover, let us impose a brickwall structure such that the circuit consists of alternating layers of two-qubit bricks.
The two types of layers are given identical probabilities for measuring the different operators: $p_{XY} = p_{YX} = sp/2$, $p_ZZ = s(1-p)$ and $p_{\mathds{1}} = 1-s$.
The transfer matrix for this circuit can be written as 
\[
    \begin{aligned}
        \hat{T} &= \hat{T}_1 \hat{T}_2 = \left( \prod_{j=1}^{\lfloor L/2 \rfloor} T_{2j-1} \right) \left( \prod_{j=1}^{\lfloor L/2 \rfloor} T_{2j} \right), \\ 
        T_l &= (1-s) \mathds{1} + \frac{sp}{2}\left(\calP_{XY} + \calP_{YX}\right) + s(1-p)\calP_{ZZ}
    \end{aligned}
\]
where $T_l$ is the transfer matrix corresponding to a brick acting on qubits $l$ and $l+1$.
The projectors $\calP_{X_l Y_{l+1}}$ and $\calP_{Y_l X_{l+1}}$ individually form the generators of a TL algebra which we denote $e_l$ and $f_l$, respectively.
For forced projections onto the $+1$ outcome, we may write $\calP_{Z_l Z_{l+1}} = \mathds{1}_{l,l+1} + 2\calP_{X_l Y_{l+1}} \calP_{Y_l X_{l+1}} - \calP_{X_l Y_{l+1}} - \calP_{Y_l X_{l+1}}$.
Averaging over the possible two-qubit states, a forced projection onto the $+1$ eigenstate of $Z_lZ_{l+1}$ followed by normalization of the wavefunction may be approximated as $\calP_{Z_l Z_{l+1}} \rightarrow 1 - e_l - f_l + 2e_lf_l$.
The transfer matrix then takes the form
\[
    T_l = (1 + s(1 - 2p)) \mathds{1} + \frac{s(3p-2)}{2}(e_l + f_l) + 2s(1-p) e_l f_l
\]
Observe that the set $\{ E_l \equiv e_l f_l \}$ generates a TL algebra.
When $p=\tfrac23$, the transfer matrix involves only this new set of generators,
\[
    T_l\vert_{p=\frac23} = \left(1 - \frac{s}{3}\right)\mathds{1} + \frac{2s}{3} E_l
\]
Since the odd and even layers of the circuit have the same measurement probabilities, this corresponds to a critical 1-state Potts model with TL algebra generators $E_l$.
Then one can identify $p_c = \tfrac23$ as a strong-coupling fixed point, in agreement with the numerical results presented in Sec.~\ref{sec:beyond_ff}.

\subsubsection{Competing Four-Fermion Measurements}

Let us now consider adding $XIX$ measurements which compete with the other four-fermion $ZZ$ measurements.
This generally spoils the previous solution for a strong coupling fixed point.
Nonetheless, suppose we now take $r_{ZZ} \neq 1$.
The circuit must now involve 3-qubit bricks, each of which has a transfer matrix like
\[
    \begin{aligned}
    T_l \sim & (1-s)\mathds{1} + \frac{sp}{2}\left(\calP_{XY} + \calP_{YX}\right) \\ 
    &+ s(1-p)\left(r \calP_{ZZ} + (1-r)\calP_{XIX}\right)
    \end{aligned}
\]
As before, we may effectively write $\calP_{X_l X_{l+2}} \sim \mathds{1} - e_l - f_{l+1} + 2e_l f_{l+1}$.
Then writing the transfer matrix in terms of the TL generators we have
\[
    \begin{aligned}
    T_l \sim &\frac{1-sp}{s} \mathds{1} + \frac{3p-2}{2}e_l + \frac{p-2(1-p)r}{2} f_l \\ 
    &- (1-p)(1-r) f_{l+1} + 2(1-p)e_l\left((1-r)f_{l+1} + r f_l\right)
    \end{aligned}
\]
As before, the coefficient on $e_l$ vanishes for $p=2/3$.
Now however there remains finite magnitude of $f_i$ and $f_{i+1}$.
Letting $p = \tfrac23 + \delta$, the two terms become $\tfrac{1-r}{3} + \delta(r+\tfrac12)$ and $\tfrac{1-r}{3} - \delta(1-r)$.
We see that these deviate from $(1-r)/3$ in opposite directions.
They take on equal and opposite magnitude at $p_c(r) = 2(1-2r)/(1 - 4r)$, which falls in the interval $[0,1]$ for $r \in [1/2, 1]$.
Since the phase diagram must be symmetric under $r \rightarrow 1-r$, up to the presence of edge modes, in the interval $r\in [0,1/2]$ we have a phase boundary at $p_c(r) = 2(1-2r)/(3-4r)$.
This is in agreement with the phase diagram obtained from numerical simulation, shown by the black dashed line in Fig.~\ref{fig:interacting_twoLeg_decoupled}c.

This argument can be made more precise by considering a full row transfer matrix, 
\[
    \begin{aligned}
    T \sim \sum_l \biggl[&\frac{s}{2}(\calP_{X_lY_{l+1}} + \calP_{Y_lX_{l+1}}) + (1-s)r\calP_{Z_lZ_{l+1}} \\ 
    &+ (1-r)\calP_{X_lX_{l+2}} \biggr]
    \end{aligned}
\]
It is now useful to take the $XY$ and $YX$ measurements to be forced projections onto $+1$ and $-1$ on opposite sublattices such that $e_{2l-1} = \calP_{X_{2l-1} Y_{2l}}$ while $e_{2l} = \mathds{1} - \calP_{X_{2l} Y_{2l+1}}$.
Then forcing $ZZ$ and $XIX$ measurements onto the $+1$ outcome gives
\[
    \begin{aligned}
    T &= \sum_l \left[ \lambda (e_l + f_l) + 2(r e_l f_l - (1-r) e_l f_{l+1})\right] \\
    &= \sum_l \left[\lambda (e_l + f_l) + e_l\left(\delta r (f_l + f_{l+1}) + (f_l - f_{l+1})\right) \right]
    \end{aligned}
\]
where we have defined $\lambda = \frac{s}{2(1-s)} - \abs{\delta r}$ and $\delta r = 2r-1$.
We see that for $s_c(r) = 2\abs{\delta r} / (1 + 2\abs{\delta r})$, the linear term vanishes (i.e. $\lambda = 0$), leaving only the couplings $e_l f_l$ and $e_l f_{l+1}$.
While we cannot construct a TL algebra from this mixture, we retain a well defined continuum limit.
Identifying $\epsilon$ and $\bar{\epsilon}$ as the energy-density fields in the CFT for $e_l$ and $f_l$, respectively, the coupling generally takes the form $\epsilon \left(2\delta r \bar{\epsilon} - \partial_x \bar{\epsilon}\right)$.
A shift in the coordinates for $\bar{\epsilon}$ then reduces the coupling to simply $2\delta r \epsilon \bar{\epsilon}$ so that the effective long-wavelength description of the circuit is simply that of the earlier interacting circuit but with a modified interaction strength.

\section{Broadening of the percolation critical point for non-Gaussian circuits}\label{sec:Broadening}

\begin{figure}[t]
    \centering
    \includegraphics[width=\columnwidth]{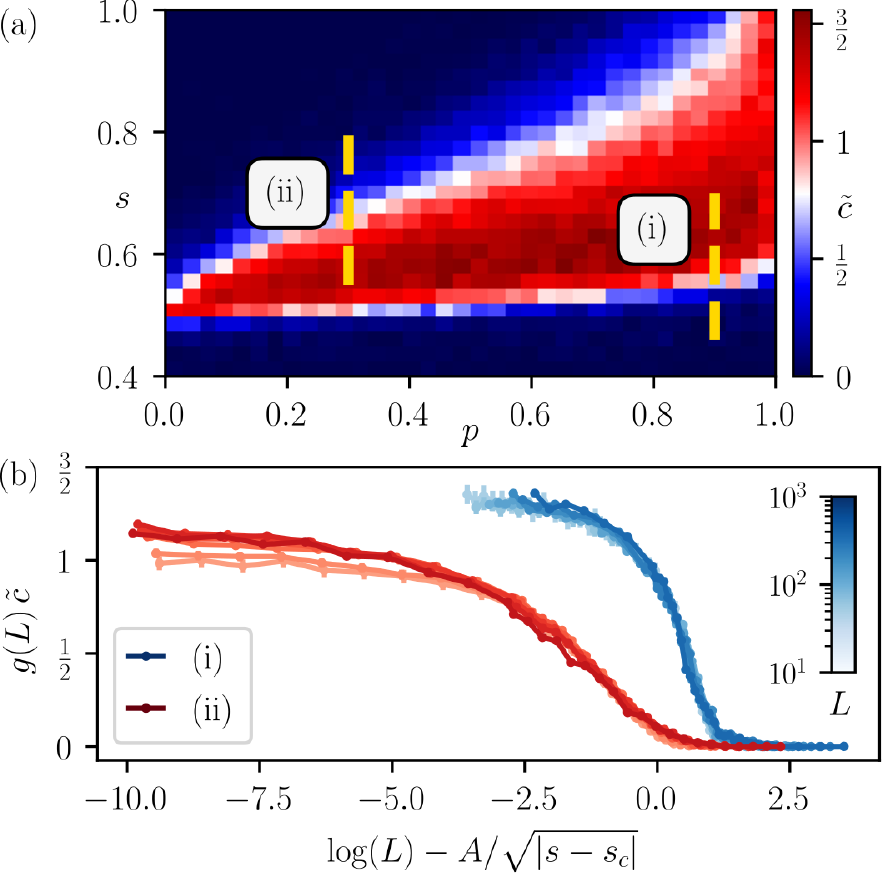}
    \caption{
    \textbf{Broadening of the percolation critical point for non-Gaussian circuits: }
    Here we take $r_{XX} = 1$, $r_{ZZ} = 1$ and $q=\tfrac12$.
    (a) Phase diagram showing the log-law coefficient $\tilde{c}$ in the $s$-$p$ plane shows an extended critical phase separating two topologically distinct area-law phases.
    The transitions terminate at an interacting percolation critical point at $(p,s) = (0,1/2)$, a decoupled critical point at $(p,s) = (1,0)$, and a BKT critical point at $(p,s) = (1,2/3)$.
    Data shown were extracted from the subsystem entanglement entropy $S_\ell$ for system size $L=256$.
    (b)
    Scaling collapse of the log-law coefficient $\tilde{c}$ from fitting the subsystem entanglement profile $S_\ell$ for the two cuts along the transition marked in (a).
    The scaling is consistent with a sudden jump in $\tilde{c}$ at a BKT transition such that $g(L)\tilde{c} = F(x)$ for $g(L) = \left[1 + \left(2\log(L) - B\right)^{-1}\right]^{-1}$ and $F$ a function of $x = \log(L) - A/\sqrt{\abs{s-s_c}}$.
    Cut (i) is taken at $p=0.9$, showing the BKT transition at $s_c \approx 0.628$ from the interacting area-law phase at $s < s_c$ to the critical phase, with fitting parameters $A \approx 1.5$ and $B \approx -9.5$.
    Cut (ii) is taken at $p=0.3$, showing the BKT transition from the Gaussian area-law phase to the critical phase at $s_c \approx 0.563$, with fitting parameter $A \approx 2.35$ and $B \approx 5.41$.
    }
    \label{fig:interacting_twoLeg_bipartite}
\end{figure}

As noted in Sec.~\ref{sec:orient_non_gauss}, when the area-law phases in the Gaussian and interacting limits of a circuit are not topologically equivalent, then there must be an entanglement transition observed upon varying the interaction strength.
Moreover, since four-fermion measurements are irrelevant in a short-loop phase, this transition ought to occur at \emph{finite} interaction strength $0 < s_c < 1$.
To see this, consider the Gaussian limit drawn from orientable subset $\mathcal{O}_3$ but excluding $YY$ measurements.
Here, $XX$ measurements drive the otherwise decoupled critical chains into an area-law phase.
Let us now perturb this circuit with orientable non-Gaussian four-fermion $ZZ$ measurements such that the interacting limit lies also in an area-law phase.
We take measurement probabilities $p_{XY} = p_{YX} = sp/2$, $p_{XX} = s(1-p)$, and $p_{ZZ} = 1-s$.
Upon varying the interaction strength, we observe an extended critical phase $s_{c,1} < s < s_{c,2}$ separating the two topologically distinct area-law phases.
The transitions from either area-law phase into the critical phase falls into the BKT universality class, as reflected in Fig.~\ref{fig:interacting_twoLeg_bipartite}b.

The apparent broadening of the transition into a critical phase can be well understood by varying not only the interaction strength but also the strength of the inter-chain $XX$ measurements which were responsible for inducing an area-law in the Gaussian limit.
As shown in Fig.~\ref{fig:interacting_twoLeg_bipartite}a, the decoupled critical point at $p=1$ is robust against finite interactions, which are RG irrelevant, giving an extended critical phase.
For weak inter-chain coupling $1-p \ll 1$, the measurement frustration with $ZZ$ measurements suppresses the effect of $XX$ measurements, effectively protecting the criticality of the decoupled limit.
In the opposing limit ($p=0$), $ZZ$ and $XX$ measurements reduce to two decoupled copies of the repetition code which undergo a percolation transition at $s=1/2$.
Here $XY$ and $YX$ measurements appear as higher-order interaction coupling between two chains at the percolation critical point, leading again to a broadening.
The now broadened critical points near $p=0,1$ extend to two BKT lines that span the two limits.
In the intermediate critical phase, the entanglement entropy has log-law coefficient $\tilde{c}\approx\tfrac32$.

\section{Ancilla Probes of Bulk Correlation Functions}\label{sec:ancilla_probe}

In order to extend the set of accessible observables from \emph{boundary} to \emph{bulk} correlations of the loop model, the knowledge of the final state at the temporal boundary is no longer sufficient and one needs to access information along the whole of the circuit trajectory. 
Here we introduce an ancilla-based detection scheme~\cite{Gullans_2020_probe}, with which $n$-particle bulk correlation functions, i.e., the watermelon correlators $G_k(x,y)$ introduced in Sec.~\ref{sec:loop_review}, can be measured in the Majorana circuit.  

\begin{figure}[t]
    \centering
    \includegraphics[width=\columnwidth]{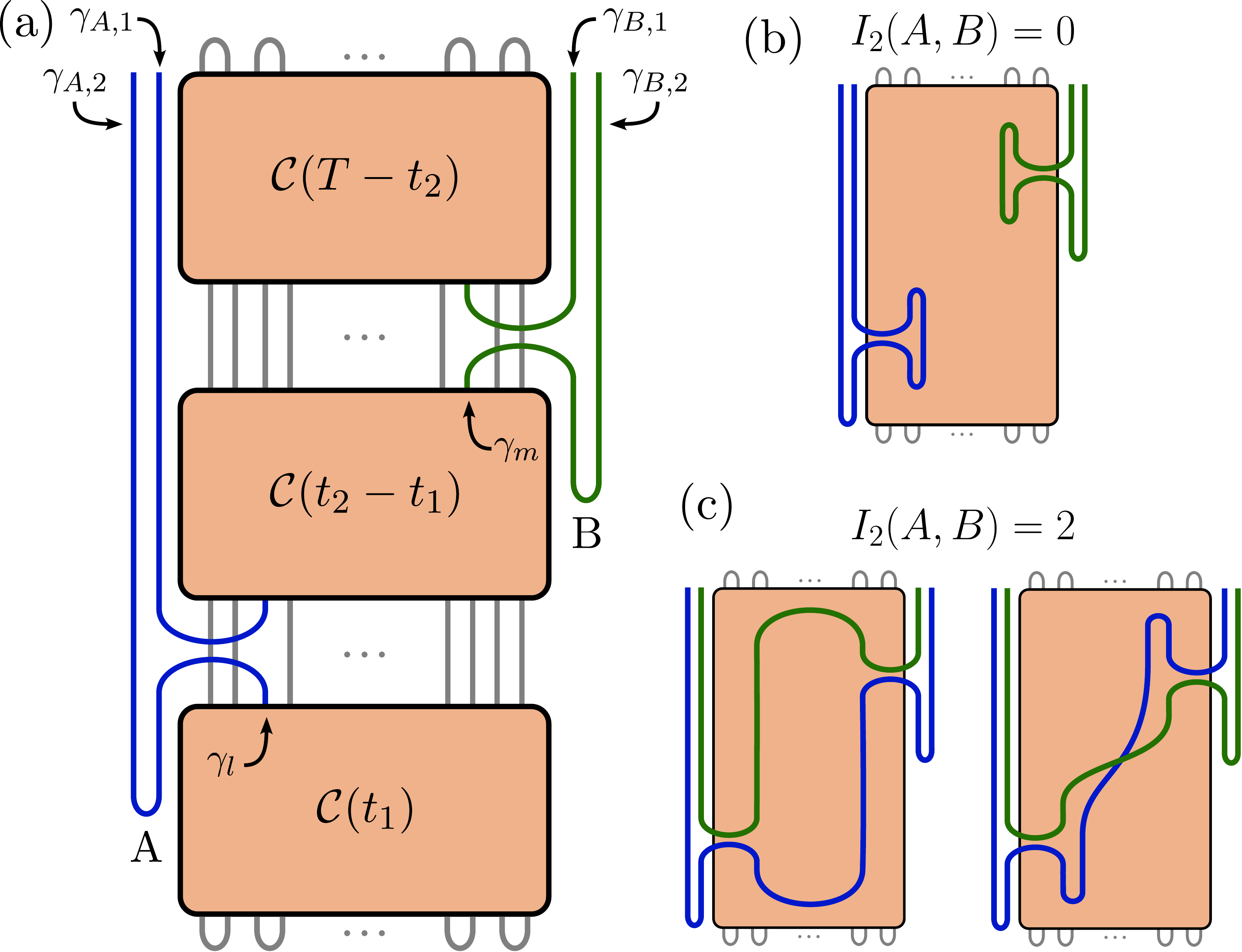}
    \caption{\textbf{Measuring bulk correlations in the circuit.}
    We present an ancilla based scheme for detecting the two-leg watermelon correlator $G_2(x,y)$ with $x = (l, t_1)$ and $y = (m, t_2)$.
    (a)
    Between times $t=0$ and $t=t_1$ an initial state $\rho_0$ is evolved by a circuit $\mathcal{C}$. At time $t_1$, a parity measurement pairs the bulk worldline $\gamma_{l}$ with worldline $\gamma_{A,1}$ from ancilla A (blue).
    The circuit then evolves to time $t_2$, when  $\gamma_{m}$ is paired with $\gamma_{B,1}$ from ancilla B (green), after which the circuit evolves to the final time $T \gtrsim L$.
    (b-c) $G_2(x,y)$ is the probability that $(x,t_1)$ and $(y, t_1)$ are on the same, closed loop, i.e., $ G_2(x,y) = \tfrac12  I_2(A,B) $ the mutual information between the ancillas. When no worldlines connect the two points (b) $I_2(A,B) = 0$ and otherwise $I_2(A,B) = 2$, see possible examples in (c).
    }
    \label{fig:watermelon_ancilla}
\end{figure}

We consider the two-leg watermelon correlator $G_2(x,y)$, which measures the probability that two points $x$ and $y$ lie upon a single, \emph{closed} loop in the two-dimensional plane. 
In the circuit, this translates to the probability that two Majorana modes $\gamma_l(t_1), \gamma_m(t_2)$ at space-time points $x=(l,t_1), y=(m,t_2)$ share a closed loop.
In order to detect $G_2(x,y)$ in the circuit, we prepare two additional pairs of ancillary worldlines, $A,B$, with $A=(\gamma_{A1}, \gamma_{A2})$ and $B=(\gamma_{B1}, \gamma_{B2})$, as depicted in Fig.~\ref{fig:watermelon_ancilla}.
Both are put in a parity eigenstate at $t=0$, $i\langle\gamma_{A1}(0) \gamma_{A2}(0)\rangle=i\langle\gamma_{B1}(0) \gamma_{B2}(0)\rangle=1$. 
Now we evolve the circuit over time but interrupt it at times $0\ll t_1,t_2\ll T$, when the parities $\gamma_{A1}(t_1)\gamma_l(t_1)$ and $\gamma_{B1}(t_2)\gamma_m(t_2)$ are measured. 
This inserts the marked loops in Fig.~\ref{fig:watermelon_ancilla} into the bulk, while storing the history of the loops crossing through spacetime points $x$ and $y$ in ancilla worldlines $\gamma_{A2}$ and $\gamma_{B2}$, respectively.

Only if the Majoranas $\gamma_l(t_1), \gamma_m(t_2)$ would share a closed loop without inserting the ancilla measurements, $A$ and $B$ will be connected via exactly two distinct worldlines. 
Otherwise no worldlines connect $A$ and $B$. 
Thus $G_2(x,y)$ gives precisely the mutual information between the two ancillas, $G_2(x,y)=\frac{1}{2}I(A,B)$.
Higher order watermelon correlators $G_k(x,y)$ can be constructed analogously by making use of further ancilla states. 
Notably, $G_2(x,y)$ appears as an out-of-time order correlator (OTOC) in the Majorana framework.
The mutual information becomes the time-nonlocal (connected) correlation function $\langle \gamma_{A1}(0)\gamma_{A1}(T) \gamma_{B1}(0)\gamma_{B1}(T)\rangle-\langle \gamma_{A1}(0)\gamma_{A1}(T)\rangle\langle \gamma_{B1}(0)\gamma_{B1}(T)\rangle$.

\section{Supplementary Data for Ising Transition}\label{app:ising_transition}

In Sec.~\ref{ss:ising} of the main text, we identified a measurement-induced transition at the tricritical point of the 1-state Potts model, for which the correlation length exponent $\nu\approx 1$ was consistent with Ising universality.
Here we provide additional characterization of the transition, examining both the critical exponents and the prefactor $\tilde{c}$ for the logarithmic entanglement scaling.

\subsection{Critical exponents $\eta$ and $\eta_\parallel$}

To further verify the Ising universality of the transition, we employ ancilla probes to extract the bulk critical exponent $\eta$ and the surface critical exponent $\eta_\parallel$.
For Ising universality realized via the 2-state Potts model we expect $\eta = \tfrac14$ and $\eta_\parallel = 1$ \cite{Cardy_1984}.
As with watermelon correlators, these exponents can be determined by examining the mutual information between ancillas which are entangled with the system\cite{Gullans_2020_probe}.
In particular, we take PBC and evolve the bulk system under measurement-only dynamics until time $t_0$.
We then maximally entangle qubit 1 with ancilla $A$ and qubit $\lfloor L/2 \rfloor + 1$ with ancilla $B$.
From the subsequent time evolution of the mutual information $I_2(t, L)$ between the ancilla, we extract exponents by scaling collapse of the form $I_2(t, L) \sim L^a F[(t-t_0)/L]$.
For $a = \eta, \eta_\parallel$ we take $t_0 = 3L, 0$, respectively.

\begin{figure}[!ht]
    \centering
    \includegraphics[width=\columnwidth]{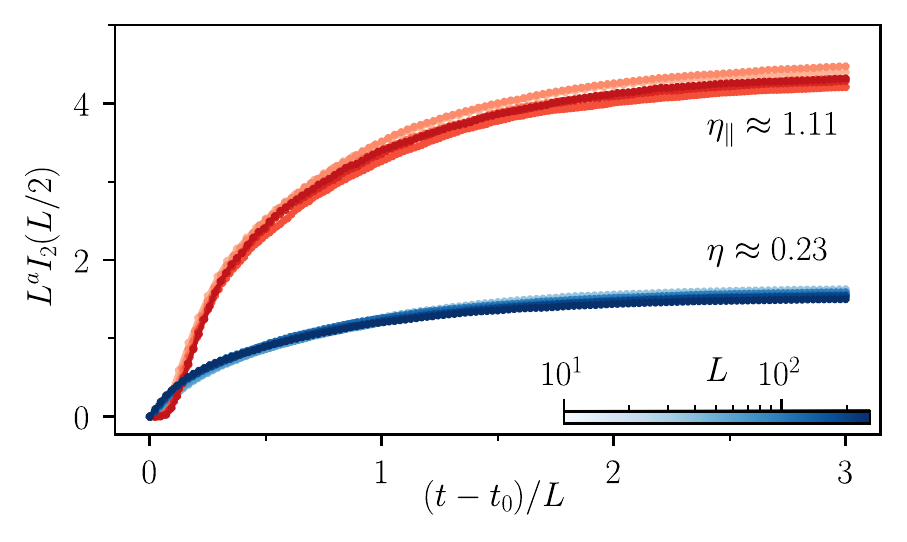}
    \caption{
    \textbf{Ancila probes for Ising criticality.}
    Evolution of the mutual information $I_2$ between ancilla coupled to the system at separation $L/2$ at time $t_0$.
    For the bulk exponent $a = \eta$ we take $t_0 = 2L$ and find $\eta\approx 0.25$, consistent with the expected $\eta=\tfrac14$ for Ising universality.
    For the surface exponent $a = \eta_\parallel$ we take $t_0 = 0$, finding $\eta_\parallel \approx 1.1$, comparable to the Ising value $\eta_\parallel = 1$.
    }
    \label{fig:ising_ancilla_evo}
\end{figure}

The rescaled mutual information is shown in Fig.~\ref{fig:ising_ancilla_evo}, from which we obtain $\eta \approx 0.23$ and $\eta_\parallel \approx 1.11$.
These are comparable to the expected values for the Ising / 2-state Potts model, with $\eta_\parallel$ being unambiguously distinct from the percolation value $\eta_\parallel^{\textrm{(perc})} = 2/3$.

The uncertainty in the trajectory averaged mutual information can be reduced by instead entangling groups of $N_A$ ancilla to the bulk system again in the regions surrounding $x=1$ and $y=\lfloor L/2 \rfloor + 1$.
As with the $N_A = 1$ case, this inserts $2N_A$ marked loops at both bulk regions.
For sufficiently large system sizes and late times, the ancilla mutual information is dominated by the largest watermelon correlator.
As such, the scaling collapse will reveal $\eta, \eta_\parallel$.
Taking $N_A > 1$ has the added advantage of eliminating possible issues where certain sites or Majoranas live in decoupled sectors, which can lead to spurious results with $N_A = 1$.
For $N_A = 2, 4$ we find similar results as for $N_A = 1$, with the estimated value for $\eta_\parallel$ more closely approaching $1$ for larger $N_A$.

\subsection{Enhanced logarithmic entanglement scaling}

\begin{figure}[!ht]
    \centering
    \includegraphics[width=\columnwidth]{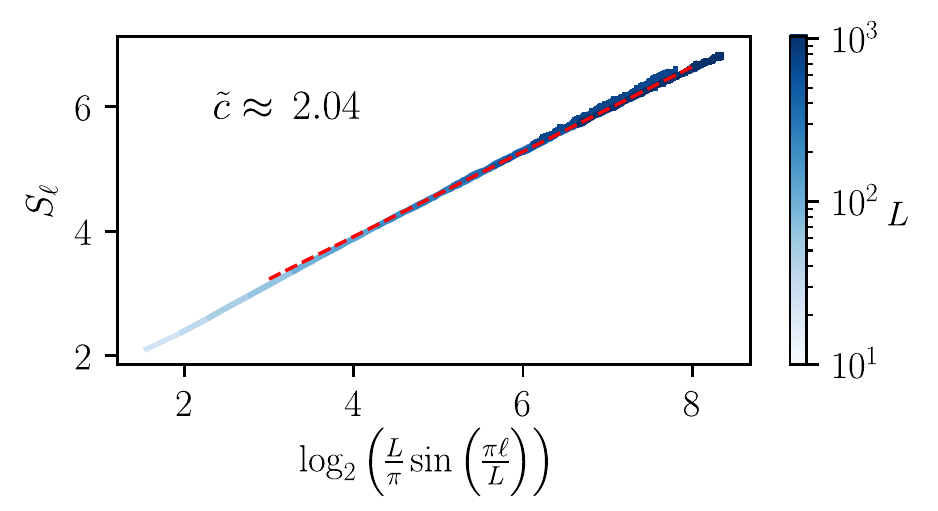}
    \caption{
    \textbf{Log-law scaling at the Ising transition.}
    At the Ising transition, the entanglement entropy $S_\ell$ for all system sizes $L$ falls onto a single line when taken against $\log_2(\sin(\pi\ell/L)L/\pi)$.
    The slope identifies the log-law coefficient $\tilde{c} \approx 2.04$ when fitted against all data for $L > 100$.
    }
    \label{fig:ising_cTilde}
\end{figure}

As noted in the main text, the coefficient $\tilde{c}$ for the log-law entanglement scaling observed at the measurement-induced Ising transition is much larger than the conventional $c=\tfrac12$.
As shown in Fig.~\ref{fig:ising_cTilde}, the subsystem entanglement entropy shows a clear log-law scaling behavior consistent with $\tilde{c} = 2$.

\section{Supplementary Data for Interaction Induced Goldstone Phase}\label{app:interaction_induced_gs}

\begin{figure}[!ht]
    \centering
    \includegraphics[width=\columnwidth]{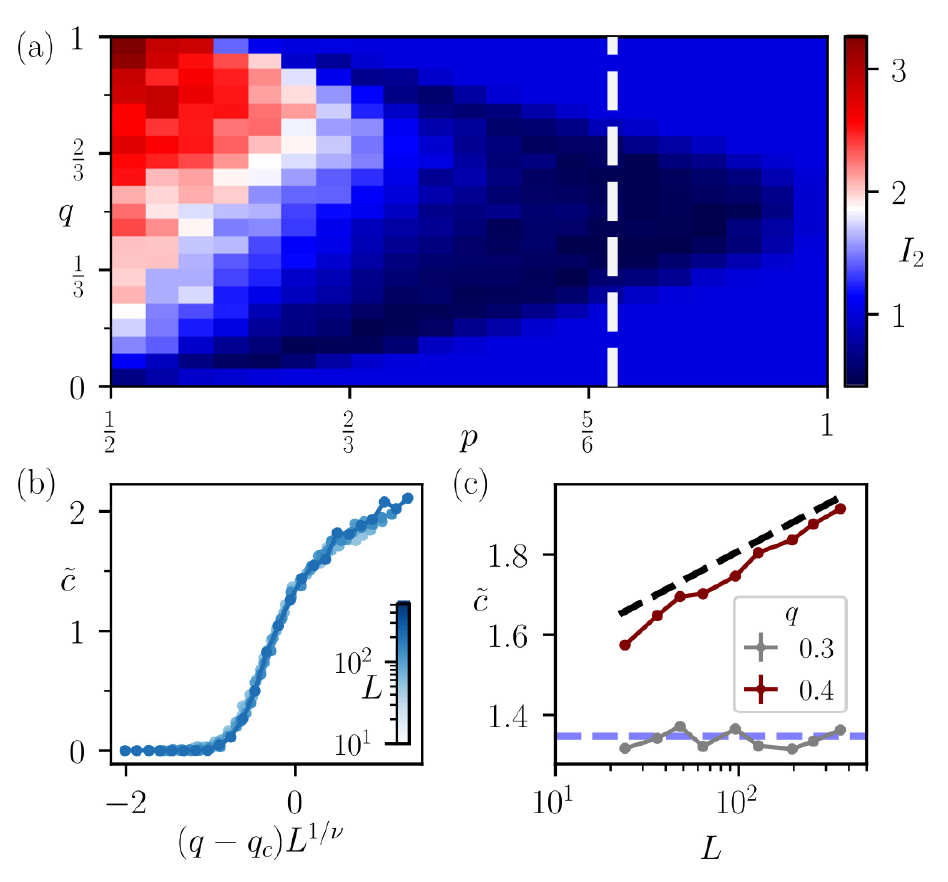}
    \caption{
    \textbf{Interaction-induced CPLC Goldstone phase.}
    Entanglement statistics in the Gaussian circuit on orientable operators $\mathcal{O}_3$ is perturbed by measurements of the non-orientable four-fermion operator $ZXY$.
    (a) Phase diagram in the $p$-$q$ plane showing an entangled phase which emerges with sufficient interaction strength and measurement frustration.
    (b) Scaling collapse of $\tilde{c}$ near the transitions along the dashed line marked in (a) reveals the critical point $q_c \approx 0.305$ and is consistent with the correlation length exponent $\nu=2.75$ expected from CPLC.
    (C) Scaling of the log-law coefficient $\tilde{c}$ with respect to system size $L$ at the transition (gray) and in the intermediate entangled phase (maroon) along the dashed line at $p=0.85$ in (a).
    The black and blue dashed lines correspond to the expected scaling of $\tilde{c}$ for the CPLC Goldstone phase and transition, respectively.
    }
    \label{fig:zxy_results}
\end{figure}

Here we provide supplementary data for the scenario discussed in Sec.~\ref{sss:interaction_induced_gs} wherein non-orientable four-fermion measurements perturbing an otherwise orientable Gaussian circuit induce a CPLC Goldstone phase.
Starting from the Gaussian circuit defined by orientable set $\mathcal{O}_3$, we introduce non-Gaussian measurements of the Pauli operator $ZXY$.
This four fermion operator is incompatible with the worldline orientations fixed by the Gaussian measurements.
Let the measurement probabilities be $p_{XX} = p_x^2$, $p_{YY} = p_y^2$, $p_{XY} = p_{YX} = p_x p_y$ and $p_{ZXY} = 1 - p$ where $p_x \equiv pq$ and $p_y \equiv p(1-q)$.
As shown in Fig.~\ref{fig:zxy_results}a, finite measurement rate for the non-orientable, non-Gaussian $ZXY$ operator yields an entangled phase fanning out from the vicinity of the maximally frustrated Gaussian point $(p,q) = (1, 1/2)$.
At sufficiently large interaction strength, the fact that $ZXY$ operators are not all mutually commuting leads to a volume-law phase, as was seen for other circuits with incommensurate four-fermion interactions.
For weaker interaction strength, the interaction-induced entangled phase resembles the CPLC Goldstone phase.
At the transitions we find correlation length exponent $\nu\approx 2.75$ and $\tilde{c}\sim\textrm{const}$, as shown in Fig.~\ref{fig:zxy_results}b,c.
Moreover, between the two critical lines we observe an apparent logarithmic correction to entanglement entropy which is consistent with CPLC (see Fig.~\ref{fig:zxy_results}c).
This provides compelling evidence that breaking worldline orientability, even if only at the 4-fermion level, leads to CPLC physics.

\end{appendix}

\end{document}